\documentclass[prb,twocolumn,showpacs,amsmath,amssymb]{revtex4}

\usepackage{graphicx}
\usepackage{dcolumn}
\usepackage{bm}
\usepackage{epsfig}

\begin{document}

\title{Decoherence of transported spin in multichannel spin-orbit coupled spintronic devices: Scattering approach to spin density matrix from the ballistic to the localized regime}

\author{Branislav K. Nikoli\' c and Satofumi Souma}
\affiliation{Department of Physics and Astronomy, University
of Delaware, Newark, DE 19716-2570, USA}

\begin{abstract}

By viewing current in the detecting lead of a spintronic device as being   
an ensemble of flowing spins corresponding to a mixed quantum state, where 
each spin itself is generally described by an improper mixture generated during 
the transport where it couples to other degrees of freedom due to spin-orbit (SO) 
interactions or inhomogeneous  magnetic fields, we introduce the {\em spin density operator} 
associated with such current expressed in terms of the spin-resolved Landauer transmission 
matrix  of the device. This formalism, which provides complete description of {\em coupled} spin-charge {\em quantum}  transport in open finite-size systems attached to external 
probes, is employed to  understand  how initially injected pure spin states, comprising 
fully spin-polarized current, evolve into the mixed ones corresponding to partially polarized 
current. We analyze particular routes that diminish spin 
coherence (signified by decay of the off-diagonal elements of the current spin density 
matrix) in two-dimensional electron gas-based devices due to the interplay of the 
Rashba and/or Dresselhaus SO coupling and: (i) scattering at the boundaries or lead-wire 
interface in ballistic semiconductor nanowires; or (ii) spin-independent scattering off 
static impurities in both weakly and  strongly localized disordered
nanowires. The physical interpretation of spin decoherence in the
course of multichannel quantum transport in terms of the {\em entanglement} of 
spin to an effectively zero-temperature ``environment'' composed of more than one 
open orbital conducting channels offers insight into some of the key challenges
for spintronics: controlling decoherence of transported spins and emergence of  
{\em partially coherent} spin states in all-electrical spin manipulation schemes based 
on the SO interactions in realistic semiconductor structures. In particular, our analysis 
elucidates why operation of both the ballistic and  non-ballistic spin-field-effect transistors, envisaged to exploit Rashba or Rashba+Dresselhaus 
SO coupling respectively, would demand single-channel transport as  the only 
set-up ensuring complete suppression of (D'yakonov-Perel'-type) spin decoherence.
\end{abstract}

\maketitle

\section{Introduction}

The major goal of recent vigorous efforts in  semiconductor spintronics is to create,  store, manipulate at a given location, and transport electron spin through conventional semiconductor  environment.~\cite{spintronics} The magneto-resistive sensors, brought about by basic research in metal  spintronics,~\cite{maekawa,prinz} have given a crucial impetus for advances in information storage technologies.  Furthermore, semiconductor-based spintronics~\cite{spintronics,spin_book} offer richer avenues for both fundamental studies and  applications because of
wider possibilities to engineer semiconductor structures by doping and gating and integrate them with conventional electronics. The two principal challenges~\cite{spin_book} for semiconductor
spintronics are: {\em spin injection}  and {\em coherent spin manipulation}.

The current efficiency of conventional spin injection into a semiconductor ($Sm$) at room 
temperature (via Ohmic contacts and at the Fermi energy), based on  ferromagnetic ($FM$) 
metallic sources of spin currents, is much lower that in the case of metal spintronic structures~\cite{jedema} due to the mismatch in the band structure and transport properties of $FM$ and $Sm$.~\cite{injection} Nevertheless,  basic transport experiments at low temperatures can evade paramount problems in spin injection 
into bulk semiconductors   by employing diluted magnetic semiconductors~\cite{ohno} or optical 
injection  techniques~\cite{optical,stevens2003} [note that spin injection and detection into 
high-mobility two-dimensional electron gas (2DEG) has turned out to be much more 
demanding~\cite{hammar}]. Also, quantum-coherent spin filters,~\cite{spin_filter} quantum spin pumps,~\cite{spin_battery} and mesoscopic generators of 
pure (i.e., not accompanied by any net charge current) spin Hall  current~\cite{meso_spin_hall} 
are expected to offer alternative solutions by making possible spin current induction without using any ferromagnetic elements. In addition, quantum-coherent spintronic devices have been proposed~\cite{ring,diego,souma}  that could make possible modulation of conventional 
(unpolarized) charge  current injected into a  semiconductor with Rashba SO interaction 
by  exploiting  spin-sensitive quantum interference effects in mesoscopic conductors of 
multiply-connected geometry (such as rings). Thus, even with successful generation of  spin currents in 
semiconductor nanostructures a challenge remains---careful manipulation of transported 
spins in classical [such as the spin-field-effect transistors (spin-FET)~\cite{datta90,nonballistic}] 
or quantum (such as  mobile spin qubits~\cite{mobile_qubit}) information  processing devices 
that will not destroy  coherent superpositions of quantum states 
$a  |\!\! \uparrow \rangle + b |\!\! \downarrow \rangle$ necessary for their operation.

The spin-FET proposal~\cite{datta90}  epitomizes one of the most
influential concepts to emerge in semiconductor spintronics---replacement of
cumbersome traditional  spin control via externally applied
magnetic fields by {\em all-electrical} tailoring of spin dynamics via 
SO interactions. Electric fields can be produced and controlled 
in far smaller volumes and on far shorter time scales than  magnetic 
fields, thereby offering  possibility for efficient local manipulation of 
spins and smooth integration with conventional high-speed digital electronic 
circuits. In the envisaged spin-FET device, spin (whose polarization  vector is 
oriented in the direction of transport) is injected from the source into 
$Sm$ wire, it precesses within this nonmagnetic region in a controlled fashion 
due to the Rashba-type~\cite{rashba} of  SO coupling (arising because of the structural 
inversion asymmetry of heterostructures) that can be tuned by the gate voltage,~\cite{nitta} 
and finally enters into the drain electrode with probability which depends on the angle of 
precession. Thus, such polarizer-analyzer electrical transport scheme would be able 
to modulate fully spin-polarized source-drain charge current.

Inasmuch as coherent spin states can be quite robust in semiconductor quantum 
wells due to weak coupling of spin to the external environment, they have  been successfully 
transported over hundreds of microns at low temperatures.~\cite{kikkawa} However, 
since SO interactions couple spin and momentum of an electron,~\cite{silsbee} they 
can also enable some of the main mechanisms leading to the decay of spin polarization~\cite{spin_book,jaro} when elastic (off lattice imperfections, nonmagnetic 
impurities, interfaces, boundaries, ...) or inelastic (off phonons)  charge scattering 
occurs in 2DEG. For example, in the semiclassical picture, put forth by D'yakonov and Perel' 
for an {\em unbounded} system with  scattering off static impurities (which does not involve  
instantaneous spin flip),~\cite{dyakonov} spin gets randomized due to  the change of the effective momentum-dependent Rashba magnetic field   ${\bf B}_{\rm R}({\bf k})$ (responsible for spin precession) in each scattering event. Thus, the  DP spin relaxation~\cite{footnote_relax} will compete with controlled Rashba spin precession, which can impede   the operation of devices involving  SO couplings. This has prompted recent  reexamination   of the spin-FET concept toward possibilities for non-ballistic modes of operation where  spins could remain coherent even in the presence of charge  scattering,~\cite{nonballistic}  in contrast to the original proposal of Datta and Das~\cite{datta90} which essentially requires clean one-dimensional wires.

While inelastic processes inevitably drive the spin polarization to zero in the long 
time limit,~\cite{inelastic} the DP spin relaxation involves only elastic scattering 
of impurities which is incapable~\cite{datta_book} of dephasing the full electron wave 
function. Therefore, in the case of quantum transport through a mesoscopic (phase-coherent) 
SO coupled $Sm$ region, where electron is  described by a single wave function,~\cite{datta_book,carlo_rmt}  the coupling between  spin polarization and charge 
currents can be interpreted as stemming from the {\em entanglement} of spin and 
orbital quantum states~\cite{peres,entangled_spin} of single electrons injected and 
detected through electrodes supporting many orbital conducting channel.~\cite{entangled_spin} Within the entangled single-particle wave function, the spin 
degree of freedom cannot be described by a pure state any more---that is, the spin becomes 
subjected to decoherence process akin to mechanisms commonly studied when open quantum systems 
becomes  entangled to usually large (and dissipative) environment.~\cite{zeh,zurek} Since 
present nanofabrication technologies yield quantum wires with more than one  open 
conducting channel at the Fermi energy (including single wall carbon nanotubes where  spin 
propagates via two channels~\cite{cnt}), it is important to quantify  the degree of coherence 
of spin transported through such structures in the presence of SO coupling.

\begin{figure}
\centerline{\psfig{file=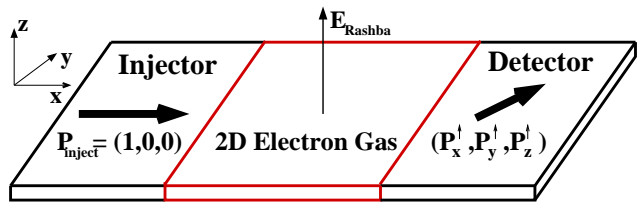,width=3.3in,angle=0} }
\caption{(Color online) Spin transport through generic two-probe spintronic device where fully spin-polarized current (comprised of pure spin states $|{\bf P}|=1$) is  injected from the left lead and detected in the right lead. The central region is 2DEG where electron can be subjected to magnetic field and/or SO interactions pertinent to semiconductor heterostructures: Rashba due to the inversion asymmetry of the confining potential; and Dresselhaus due to the bulk inversion asymmetry. If the injected current is fully spin polarized, such as  along the $x$-axis $(P_x=1,P_y=0,P_z=0)$ chosen in the Figure, the outgoing current will, in general, have its polarization vector  rotated by coherent spin precession in semiconductor region, as well as shrunk $|{\bf P}|<1$ due to processes which lead to loss of spin quantum coherence (such as spin-independent scattering at static impurities or interfaces in the presence of SO coupling).} \label{fig:setup}
\end{figure}

To loss of coherence~\cite{zeh,zurek} of transported spins is encoded into the decay 
of the off-diagonal elements of their density matrix $\hat{\rho}_s$. Recent theoretical pursuits 
have offered diverse approaches~\cite{burkov,wu,privman,cahay,mischenko,spin_ballistic,sinova_bloch}  that make it possible to follow the quantum dynamics of $\hat{\rho}_s$ in the course of transport, while treating the  ballistic~\cite{mischenko,spin_ballistic} or the diffusive~\cite{burkov,privman,cahay}  propagation of charges to which the spins are attached {\em semiclassically}. The Landauer-B\" uttiker scattering formalism,~\cite{datta_book,carlo_rmt} which intrinsically takes into account phase-coherent propagation of electrons through finite-size devices 
attached to external current and voltage probes, is also frequently employed to treat quantum 
spintronic transport in semiconductor structures.~\cite{mireles,feve,pareek,tang,popp} However, 
previous applications of the scattering formalism evaluate only the spin-resolved charge 
conductances  which, on the other hand, do not provide enough information to extract the 
full density matrix of transported spins, ``hiding'' in the quantum transmission properties of 
the device. Such approaches provide only a single component of the spin polarization vector of detected current in the right lead of Fig.~\ref{fig:setup}, while all three 
components are needed to: (i) determine the vector of spin current flowing together with 
charge current in this lead; (ii) evaluate the density matrix of the corresponding ensemble of 
transported spins; and (iii) extract their degree of coherence.~\cite{zeh,zurek,galindo}

Here we demonstrate how to associate the spin density matrix to 
detected current, which emerges after charge current with arbitrary 
spin-polarization properties (unpolarized, partially polarized, or fully spin-polarized) 
is injected through multichannel leads and propagated through 
quantum-coherent semiconductor nanostructure where transported electrons are 
subjected to spin-dependent interactions.  Following our earlier analysis of the 
density matrix of a single spin injected through one of the Landauer conducting channels,~\cite{entangled_spin} we introduce in Sec.~\ref{sec:purity} a density 
matrix of an ensemble of spins flowing through the detecting lead in 
Fig.~\ref{fig:setup}. This central tool of our approach is expressed in terms of 
both the amplitudes and the phases of (spin-resolved) Landauer transmission 
matrix elements. In Sec.~\ref{sec:polarization} we extract from it the spin polarization 
vector $(P_x^\sigma,P_y^\sigma,P_z^\sigma)$ of the outgoing current in Fig.~\ref{fig:setup} 
while taking into account different possibilities for the polarization $\sigma$ of the 
incoming current. This also allows us to elucidate rigorous way of 
quantifying the spin polarization (as a scalar quantity) of current which is measured 
in spin detection experiments.~\cite{hammar,silsbee,marcus} Together with the Landauer 
formulas for spin-resolved charge conductances (that involve only the squared 
amplitudes of the transmission matrix elements~\cite{mireles,feve,pareek,tang,popp}), 
our equations for $(P_x^\sigma,P_y^\sigma,P_z^\sigma)$ offer complete  description of 
the coupled spin-charge quantum transport in finite-size devices where experimentally 
relevant boundary  conditions (such as closed boundaries at which current must vanish, 
interfaces, external electrodes, and spin-polarization properties of  the injected 
current), which are crucial for the treatment of transport in the presence of SO couplings, 
are easily incorporated.

The magnitude of ${\bf P}$ quantifies the degree of coherence of the spin state. We employ this 
formalism in Sec.~\ref{sec:decoherence} to study how spin-orbit entanglement affects transport, 
entailing the reduction of $|{\bf P}|$ in ballistic (Sec.~\ref{sec:ballistic}) or disordered (Sec.~\ref{sec:diffusive}) semiconductor multichannel quantum wires. This also offers a direct 
insight into the dynamics of quantum coherence of spin which would propagate through 
multichannel ballistic~\cite{datta90} (with Rashba coupling) or non-ballistic 
(with Rashba=Dresselhaus coupling) spin-FET~\cite{nonballistic} devices. For 
the transport of non-interacting electrons through finite-size structures, 
$(P_x^\sigma,P_y^\sigma,P_z^\sigma)$ can be evaluated non-perturbatively in both the SO couplings 
and the disorder strength. This makes it possible to treat the dynamics 
of spin coherence in a wide range of transport  regimes (from high mobility in ballistic to low mobility in localized systems), thereby unearthing quantum effects in the evolution of $|{\bf P}|$ 
that  go beyond conventional  semiclassical~\cite{dyakonov} or perturbative 
quantum treatments~\cite{wl_spin} of spin relaxation in diffusive bulk 
semiconductors with weak SO interaction. We conclude in Sec.~\ref{sec:conclusion} 
by highlighting requirements to combat spin decoherence in spintronic devices relying on 
fully coherent spin states, while also pointing out at capabilities of 
{\em partially} coherent spin states that inevitably emerge in multichannel 
devices examined here.

\section{Purity of transported spin states} \label{sec:purity}

For the understanding of quantum dynamics of open spin system and processes which leak its coherence into the environment,~\cite{zeh,zurek} the central role is played by the density operator~\cite{galindo,ballentine} $\hat{\rho}_s$. The expectation value $\langle \Sigma | \hat{\rho}_s | \Sigma \rangle$ gives the probability of observing the system in state $| \Sigma \rangle$. For spin-$\frac{1}{2}$ particle, this operator has a simple representation in a chosen basis~\cite{ballentine} $|\!\! \uparrow \rangle, |\!\! \downarrow \rangle \in {\mathcal H}_s$,
\begin{equation} \label{eq:density_matrix}
\hat{\rho}_s=\left( \begin{array}{cc}
        \rho_{\uparrow\uparrow} & \rho_{\uparrow\downarrow} \\
        \rho_{\downarrow\uparrow} & \rho_{\downarrow\downarrow}
         \end{array} \right)= \frac{\hat{I}_s+ {\bf P} \cdot \hat{\bm{\sigma}}}{2}, 
\end{equation}
which is a $2 \times 2$ spin density matrix  where $\hat{I}_s$ is the unit operator in 
the spin Hilbert space and $\hat{\bm{\sigma}}=(\hat{\sigma}_x,\hat{\sigma}_y,\hat{\sigma}_z)$ 
is the vector of the Pauli spin matrices. The diagonal elements $\rho_{\uparrow\uparrow}$ and $\rho_{\downarrow\downarrow}$ represent the probabilities to find electron with spin-$\uparrow$ or spin-$\downarrow$. The off-diagonal elements $\rho_{\uparrow\downarrow}$, $\rho_{\downarrow\uparrow}$ define the amount by which the probabilities of coherent superpositions of basis vectors $|\!\! \uparrow \rangle, |\!\! \downarrow \rangle$ deviate,  due to quantum interference effects, from the classical (incoherent) mixture of states. The two-level system density  matrix Eq.~(\ref{eq:density_matrix}) is the simplest 
example of its kind since it is determined just by a set of three real numbers 
representing the components of the spin polarization~\cite{galindo,ballentine} (or Bloch) vector 
${\bf P}=(P_x,P_y,P_z)$. For spin-$\frac{1}{2}$ particles, the polarization vector is experimentally 
measured as the quantum-mechanical average
\begin{equation}\label{eq:pol_vector}
\frac{\hbar}{2} {\bf P}=\frac{\hbar}{2}(\langle \hat{\sigma}_x \rangle,\langle
\hat{\sigma}_y \rangle,\langle \hat{\sigma}_z  \rangle ) = {\rm Tr}\,
      \left[ \hat{\rho}_s \frac{\hbar}{2} \hat{\bm{\sigma}} \right],
\end{equation}
which is the expectation value of the spin operator $\hbar \hat{\bm{\sigma}}/2$.

A fully coherent state of spin-$\frac{1}{2}$ particle is {\em pure} and, therefore, described formally by a  vector $|\Sigma \rangle$ belonging to the two-dimensional Hilbert space $|\Sigma \rangle \in H_{s}$. The density operator formalism encompasses both {\em pure} $\hat{\rho}=|\Sigma \rangle \langle \Sigma|$ states  and {\em mixtures} $\hat{\rho} = \sum_i w_i |\Sigma_i \rangle \langle \Sigma_i|$ describing an ensemble of quantum states appearing with different classical probabilities $w_i$. One can quantify the degree of coherence of a quantum state~\cite{zeh} by the purity ${\mathcal P} = {\rm Tr} \, \hat{\rho}^2$. However, since the density operator $\hat{\rho}_s$ of a spin-$\frac{1}{2}$ particle is determined solely by  the polarization vector  ${\bf P}$,  all relevant information about its coherence can be contained from the magnitude $|{\bf P}| = \sqrt {P_x^2 + P_y^2 + P_z^2}$, so that ${\mathcal P}_s = (1+|{\bf P}|^2)/2$ (note that in the case of, 
e.g., spin-1 particle one has to measure additional five parameters~\cite{ballentine} to 
specify $\hat{\rho}_s$ and its purity).

For fully coherent pure states the polarization vector has unit magnitude $|{\bf P}|=1$, 
while $0 \le |{\bf P}| < 1$ accounts for mixtures. The dynamics of electron spin is affected  
by external magnetic field, local magnetic fields produced by magnetic impurities and nuclei, 
and different types of SO couplings. These interactions not only generate quantum-coherent 
evolution of the carrier spin, but can also induce spin decoherence.~\cite{spin_book,zeh,zurek} 
Thus,  coherent motion is encoded into the rotation of vector ${\bf P}$, while 
the decay of spin coherence is measured by the reduction of its magnitude 
$|{\bf P}|$ below one. Figure~\ref{fig:setup} illustrates how these generic features in 
the dynamics of open two-level systems will manifest for spins in non-equilibrium steady 
transport state. 

\subsection{Spin density matrix of detected current} \label{sec:density_matrix}

Most of the traditional mesoscopic experiments~\cite{webb} explore superpositions of orbital states of 
transported spin-degenerate  electrons since inelastic dephasing processes are suppressed in small enough structures ($L < 1\mu$m) at low temperatures ($T \ll 1$K). This means that electron is  described by a single orbital wave function   $|\Psi \rangle \in {\mathcal H}_o$ within the conductor.~\cite{datta_book,carlo_rmt} When spin-polarized electron is injected into 
a phase-coherent semiconductor structures where it becomes subjected to interactions 
with effective magnetic fields, its state will remain pure, but now in the tensor product 
of the orbital and the spin Hilbert spaces $|\Psi \rangle \in {\mathcal H}_o \otimes {\mathcal H}_s$. 
Inside  the ideal (free from spin and charge interactions) leads attached to the sample, electron wave function can be expressed as a linear combination of spin-polarized conducting channels $|n \sigma \rangle = |n \rangle \otimes |\sigma\rangle$ at a given Fermi energy. Each channel, being a tensor product of the orbital transverse propagating mode and a spinor, is a separable~\cite{galindo} pure quantum state $\langle {\bf r} |n \sigma \rangle^\pm =  \Phi_n(y) \otimes \exp(\pm i k_n x) \otimes | \sigma \rangle$ specified by a real wave number $k_n > 0$, transverse mode $\Phi_n(y)$ defined by the quantization of transverse momentum in the leads of a finite cross section, and a spin  factor state $|\sigma \rangle$ (we assume that orbital channels are normalized in a usual way to carry a unit current~\cite{carlo_rmt}). When injected spin-polarized flux from the left lead of a two-probe device is concentrated in the spin-polarized channel $|{\rm in}\rangle \equiv | n \sigma \rangle$, a pure state emerging in the right lead will, in general, be described by the linear combination of the outgoing channels
\begin{equation} \label{eq:flux}
|{\rm out} \rangle = \sum_{n^\prime\sigma^\prime} {\bf t}_{n^\prime n,\sigma^\prime
\sigma} |n^\prime \rangle \otimes |\sigma^\prime \rangle,
\end{equation}
which is a {\em non-separable}~\cite{galindo} state.
This equation introduces the spin-resolved Landauer transmission matrix: $|{\bf t}_{n^\prime
n,\sigma^\prime \sigma}|^2$ represents the probability for a spin-$\sigma$ electron incoming from the left lead in the orbital state $|n \rangle$ to appear as a spin-$\sigma^\prime$ electron in the orbital channel $|n^\prime \rangle$ in the right lead. The matrix elements of ${\bf t}$ depend on the Fermi energy $E_F$ at which quantum (i.e., effectively zero temperature) transport takes place. The ${\bf t}$-matrix, extended to include the spin degree of freedom and spin-dependent single-particle interactions in quantum transport,~\cite{mireles,feve} is a standard tool to obtain the spin-resolved conductances of a two-probe device
\begin{equation}\label{eq:conductance}
{\bf G}  =  \left( \begin{array}{cc}
     G^{\uparrow\uparrow} &  G^{\uparrow\downarrow} \\
      G^{\downarrow\uparrow} &  G^{\downarrow\downarrow}
  \end{array} \right) =   \frac{e^2}{h} \, \sum_{n^\prime,n=1}^M
     \left( \begin{array}{cc}
      |{\bf t}_{n^\prime n,\uparrow\uparrow}|^2 &  |{\bf t}_{n^\prime n,\uparrow\downarrow}|^2 \\
      |{\bf t}_{n^\prime n,\downarrow\uparrow}|^2 & |{\bf
     t}_{n^\prime n,\downarrow\downarrow}|^2 \end{array} \right).
\end{equation}
Here $M$ is the number of orbital conducting channels (the number of spin-polarized  conducting
channels  is $2M$) determined by the properties of transverse confining potential in the leads. In the
Landauer picture of spatial separation of single-particle coherent and many-body inelastic processes,~\cite{landauer} it is assumed that sample is attached to huge electron reservoirs with negligible spin-dependent interactions. To simplify the scattering boundary conditions, semi-infinite ideal leads  are inserted between the reservoirs (which thermalize electrons and ensure steady-state transport) and the semiconductor region.

Selecting the spin-resolved elements of ${\bf t}$-matrix (see Sec.~\ref{sec:decoherence}) allows 
one to describe different spin injection and detection transport measurements. That is, the 
spin-resolved conductances can be interpreted as describing injection, transport, and 
detection of single spin-species in a set-up involving spin filters or half-metallic ferromagnetic 
leads. For example, $G^{\uparrow\downarrow}$ is the conductance of a set-up where spin-$\downarrow$ current is injected and spin-$\uparrow$ is detected. If both spin species are injected from the left lead in equal proportion, as in usual experiments with conventional unpolarized current, one resorts to the usual  Landauer conductance formula~\cite{datta_book,carlo_rmt} $G=G^{\uparrow\uparrow} + G^{\uparrow\downarrow}  + G^{\downarrow\uparrow} + G^{\downarrow \downarrow}$.

While the conductance formulas Eq.~(\ref{eq:conductance}) require to evaluate only the amplitude of the
${\bf t}$-matrix elements, Eq.~(\ref{eq:flux}) reveals that both the amplitude and the phase of ${\bf t}_{n^\prime n,\sigma^\prime\sigma}$ determine the non-separable electron state in the outgoing lead. Although $|{\rm out} \rangle$ state Eq.~(\ref{eq:flux}) is still a pure one, spin in such state is entangled to orbital conducting channels, i.e., it cannot be assigned a single spinor wave function 
as in the case of $|{\rm in}\rangle$. Obviously, such SO entanglement will be generated  whenever  orbital and spin part of the Hamiltonian do not commute, as in  the cases where, e.g., inhomogeneous  magnetic field,~\cite{popp} random magnetic impurities, or SO interaction term+inhomogeneous 
spatial potential~\cite{lyanda} govern quantum evolution of the system.

To each of the outgoing pure states of Eq.~(\ref{eq:flux}), we associate a density matrix
$\hat{\rho}=|{\rm out} \rangle \langle {\rm out}|$
\begin{equation}
\hat{\rho}^{n \sigma \rightarrow {\rm out}}= \frac{1}{Z} \sum_{n^\prime
n^{\prime\prime} \sigma^\prime \sigma^{\prime\prime}} {\bf
t}_{n^\prime n, \sigma^\prime \sigma} {\bf t}_{n^{\prime
\prime} n, \sigma^{\prime\prime} \sigma}^* |n^\prime \rangle \langle
n^{\prime\prime}| \otimes |\sigma^\prime \rangle \langle
\sigma^{\prime\prime} |,
\end{equation}
where $Z$ is a normalization factor ensuring that ${\rm Tr} \, \hat{\rho}=1$.
After taking the partial trace~\cite{zurek,ballentine} over the orbital degrees of freedom, which amounts to
summing all $2 \times 2$ block matrices along the diagonal of $\hat{\rho}^{n \rightarrow {\rm out}}$,
we arrive at the density matrix describing the quantum state of spin in the right lead.~\cite{entangled_spin}
For example, when spin-$\uparrow$ electron is injected in channel $|n \rangle$ from the left lead, 
the incoming state is $|n \rangle \otimes |\!\! \uparrow \rangle$ and the explicit form of the  density matrix for the outgoing spin state in the right lead is given by 
\begin{equation} \label{eq:rightlead}
\hat{\rho}^{n \! \uparrow \rightarrow {\rm out}}_s = \frac{1}{Z} \sum_{n^\prime =1}^M
\left( \begin{array}{cc}
     |{\bf t}_{n^\prime n,\uparrow \uparrow}|^2 &  {\bf t}_{n^\prime n,\uparrow
       \uparrow} {\bf t}_{n^\prime n,\downarrow \uparrow}^* \\
      {\bf t}_{n^\prime n,\uparrow\uparrow}^* {\bf t}_{n^\prime n,\downarrow\uparrow}
      &  |{\bf t}_{n^\prime n,\downarrow\uparrow}|^2
  \end{array} \right).
\end{equation}
Since the full outgoing state Eq.~(\ref{eq:flux}) of an electron is still pure, the reduced density matrix
$\hat{\rho}^{n \sigma \rightarrow {\rm out}}_s$ does not correspond to any real ensemble of quantum
states (i.e., it is an {\em improper} mixture~\cite{zeh}). On the other hand, the current can 
be viewed as a real ensemble of electrons injected in different channels, so that we consider spin and 
charge flow in the right lead to give rise to an ensemble of states described by a {\em proper} mixture   $\hat{\rho}_c = \sum_n  \hat{\rho}^{n \sigma \rightarrow {\rm out}}$. Thus, when spin-$\uparrow$ polarized current is injected from the left lead, we obtain for the current spin density matrix in the right lead
\begin{equation} \label{eq:rightleadup}
\hat{\rho}^\uparrow_c = \frac{e^2/h}{G^{\uparrow \uparrow} + G^{\downarrow \uparrow}} \sum_{n^\prime,n=1}^M
\left( \begin{array}{cc}
     |{\bf t}_{n^\prime n,\uparrow \uparrow}|^2 &  {\bf t}_{n^\prime n, \uparrow \uparrow}
       {\bf t}^*_{n^\prime n,\downarrow \uparrow}  \\
      {\bf t}^*_{n^\prime n,\uparrow \uparrow} {\bf
    t}_{n^\prime n,\downarrow \uparrow} &  |{\bf t}_{n^\prime n,\downarrow\uparrow}|^2
  \end{array} \right).
\end{equation}
By the same token, the spin density matrix of the detected current, emerging after the injection 
of  spin-$\downarrow$ polarized charge current, is given by 
\begin{equation} \label{eq:rightleaddown}
\hat{\rho}^\downarrow_c = \frac{e^2/h}{G^{\uparrow \downarrow} + G^{\downarrow \downarrow}} \sum_{n^\prime,n=1}^M
\left( \begin{array}{cc}
     |{\bf t}_{n^\prime n,\uparrow \downarrow}|^2 &  {\bf t}_{n^\prime n,\uparrow \downarrow}
       {\bf t}^*_{n^\prime n,\downarrow \downarrow}  \\
      {\bf t}^*_{n^\prime n,\uparrow \downarrow} {\bf t}_{n^\prime n,\downarrow \downarrow} &
    |{\bf t}_{n^\prime n,\downarrow \downarrow}|^2
  \end{array} \right).
\end{equation}
The most general case is obtained after the injection of partially spin-polarized current, whose spins 
are in the mixed quantum state
\begin{equation} \label{eq:incoherent_rho}
\hat{\rho}_s  = n_\uparrow |\!\! \uparrow\rangle \langle \uparrow \!\!| + n_\downarrow |\!\!\downarrow \rangle \langle \downarrow \!\! |,
\end{equation}
which gives rise to the following spin density matrix of the outgoing current
\begin{widetext}
\begin{equation} \label{eq:rightleadunpol}
\hat{\rho}^{\uparrow + \downarrow}_c =  \frac{e^2/h}{n_\uparrow (G^{\uparrow \uparrow} + G^{\downarrow \uparrow}) + n_\downarrow(G^{\uparrow \downarrow} +
G^{\downarrow \downarrow})} \sum_{n^\prime,n=1}^M
\left( \begin{array}{cc}
     n_\uparrow |{\bf t}_{n^\prime n,\uparrow \uparrow}|^2 + n_\downarrow |{\bf t}_{n^\prime n,\uparrow \downarrow}|^2  &
       n_\uparrow {\bf t}_{n^\prime n, \uparrow \uparrow}
       {\bf t}^*_{n^\prime n,\downarrow \uparrow} + n_\downarrow {\bf t}_{n^\prime n,\uparrow \downarrow}
       {\bf t}^*_{n^\prime n,\downarrow \downarrow}  \\
         n_\uparrow {\bf t}^*_{n^\prime n,\uparrow \uparrow} {\bf
    t}_{n^\prime n,\downarrow \uparrow} + n_\downarrow {\bf t}^*_{n^\prime n,\uparrow \downarrow} {\bf t}_{n^\prime n,\downarrow \downarrow} &
       n_\uparrow |{\bf t}_{n^\prime n,\downarrow\uparrow}|^2 + n_\downarrow |{\bf t}_{n^\prime n,\downarrow \downarrow}|^2
  \end{array} \right).
\end{equation}
\end{widetext}
This density matrix reduces to Eq.~(\ref{eq:rightleadup}) or  Eq.~(\ref{eq:rightleaddown}) in the limits $n_\uparrow=1$, $n_\downarrow=0$ or $n_\uparrow=0$, $n_\downarrow=1$, respectively.

The measurement of any observable quantity $O_s$ on the spin subsystem within the right lead is described by the reduced spin density matrix $\langle O_s \rangle = {\rm Tr}_s \, [\hat{\rho}_c \hat{O}_s$], where $\hat{O}_s$ is a Hermitian operator acting solely in  ${\mathcal H}_s$. An example of such measurement is the spin operator itself in  Eq.~(\ref{eq:pol_vector}). In the case of semiconductor quantum wires explored in Sec.~\ref{sec:ballistic} and Sec.~\ref{sec:diffusive}, the spin density matrices in Eqs. ({\ref{eq:rightleadup})--({\ref{eq:rightleadunpol}) are determined by the polarization of injected current, number of orbital conducting channels in the leads, and spin and charge-dependent interactions within the wire. They characterize transported electron spin in an open  quantum system, and can be easily generalized to multi-probe geometry for samples attached to more than two leads.

\subsection{Spin-polarization of charge currents in semiconductor spintronics} \label{sec:polarization}

What is the spin polarization of current flowing through a spintronic device? In many  metal and
insulator spintronic structures,~\cite{jedema,prinz} as well as in some of the semiconductor ones,~\cite{spin_battery} spin-up  $I^\uparrow$ and spin-down currents $I^\downarrow$ comprising charge current $I=I^\uparrow + I^\downarrow$ are independent of each other and spin quantization axis 
is usually well-defined by external magnetic fields. Therefore, spin-polarization is easily 
quantified by a single number~\cite{jedema,maekawa,prinz}
\begin{equation}\label{eq:polarization}
P=\frac{I^\uparrow - I^\downarrow}{I^\uparrow + I^\downarrow} = \frac{G^{\uparrow \uparrow} - G^{\downarrow \downarrow}}{G^{\uparrow \uparrow} + G^{\downarrow \downarrow}}.
\end{equation}
Using the language of spin density matrices,  partially polarized current $P \neq 0$ is incoherent statistical mixture of $|\!\! \uparrow \rangle$ and $|\!\! \downarrow \rangle$ states described by  Eq.~(\ref{eq:incoherent_rho}) [for $n_\uparrow=n_\downarrow$ we get the conventional completely unpolarized charge current $\hat{\rho}_s=\hat{I}_s/2 \Rightarrow |{\bf P}| =0$].

Surprisingly enough, quite a few apparently different quantities
have been proposed in recent spintronic literature to quantify the
spin polarization of detected current in semiconductor
devices.~\cite{pareek,seba,strong_rashba,popp} In semiconductors
with SO coupling, or spatially dependent interaction with
surrounding spins and external inhomogeneous magnetic fields,~\cite{popp} 
a non-zero off-diagonal spin-resolved conductances $G^{\uparrow
\downarrow} \neq 0 \neq G^{\downarrow \uparrow}$ will emerge due
to spin precession or instantaneous spin-flip processes. Thus, in contrast
to Eq.~(\ref{eq:polarization}), these expressions~\cite{pareek,seba,strong_rashba,popp} 
for ``spin polarization'' involve all four spin-resolved conductances defined by Eq.~(\ref{eq:conductance}). However, they effectively evaluate just one component of the spin 
polarization vector along the spin quantization axis (which is 
usually fixed by the direction of magnetization of ferromagnetic 
elements or axis of spin filter which specify the orientation of 
injected spins in Fig.~\ref{fig:setup}). For example, standard applications of the Landauer-B\" uttiker scattering formalism to ballistic~\cite{tang} or diffusive transport in 2DEG with Rashba SO interaction,~\cite{pareek} where only spin-resolved charge conductances are evaluated through Eq.~(\ref{eq:conductance}), allows one to obtain only $P_x^\uparrow$ in the right lead in Fig.~\ref{fig:setup}. The knowledge of  $P_x^\uparrow$ alone is insufficient to quantify the quantum coherence properties of detected spins. Also, in the case of transport of fully coherent spins, 
where $|{\bf P}|=1$ in the right lead, we need to know all three components of the outgoing polarization  vector to understand different transformations  that the device can perform on 
the incoming spin.~\cite{mobile_qubit,diego,souma} 

Our formalism provides direct algorithm to obtain the explicit 
formulas for $(P_x^\sigma,P_y^\sigma,P_z^\sigma)$  from the  spin density matrix Eq.~(\ref{eq:rightleadunpol}) by evaluating the expectation value of the spin 
operator in Eq.~(\ref{eq:pol_vector}). When injected 
current through the left lead is spin-$\uparrow$ polarized, the spin polarization vector of the
current in the right lead is obtained from
Eq.~(\ref{eq:pol_vector}) and Eq.~(\ref{eq:rightleadup}) as
\begin{subequations} \label{eq:p_up}
\begin{eqnarray}
P^\uparrow_x &  = &  \frac{G^{\uparrow \uparrow} - G^{\downarrow \uparrow}}{G^{\uparrow \uparrow} + G^{\downarrow \uparrow}}, \\
P^\uparrow_y & = & \frac{2e^2/h}{G^{\uparrow \uparrow} + G^{\downarrow \uparrow}} \sum_{n^\prime,n=1}^M {\rm Re} \,
\left[{\bf t}_{n^\prime n, \uparrow \uparrow} {\bf t}^*_{n^\prime n,\downarrow \uparrow} \right], \\
P^\uparrow_z & = & \frac{2e^2/h}{G^{\uparrow \uparrow} + G^{\downarrow \uparrow}} \sum_{n^\prime,n=1}^M {\rm Im} \,
\left[{\bf t}^*_{n^\prime n, \uparrow \uparrow} {\bf t}_{n^\prime n,\downarrow \uparrow} \right].
\end{eqnarray}
\end{subequations}
Here, and in the formulas below, the $x$-axis is chosen arbitrarily as the spin quantization axis (Fig.~\ref{fig:setup}), $\hat{\sigma}_x  |\!\! \uparrow \rangle = + |\!\! \uparrow \rangle$ and $\hat{\sigma}_x |\!\! \downarrow \rangle=  - |\!\! \downarrow \rangle$, so that Pauli spin algebra has the following representation
\begin{equation}\label{eq:pauli}
 \hat{\sigma}_x=  \left( \begin{array}{cc}
       1 & 0 \\
       0 & -1
  \end{array} \right),
   \hat{\sigma}_y = \left( \begin{array}{cc}
       0 & 1 \\
       1 & 0
  \end{array} \right),
    \hat{\sigma}_z = \left( \begin{array}{cc}
       0 & - i \\
       i & 0
  \end{array} \right).
\end{equation}
Analogously, if the injected current is 100\% spin-$\downarrow$ polarized along the $x$-axis we get 
\begin{subequations} \label{eq:p_down}
\begin{eqnarray}
P^\downarrow_x &  = &  \frac{G^{\uparrow \downarrow} - G^{\downarrow \downarrow}}{G^{\uparrow \downarrow} + G^{\downarrow \downarrow}}, \\
P^\downarrow_y & = & \frac{2e^2/h}{G^{\uparrow \downarrow} + G^{\downarrow \downarrow}} \sum_{n^\prime,n=1}^M {\rm Re} \,
\left[{\bf t}_{n^\prime n, \uparrow \downarrow} {\bf t}^*_{n^\prime n,\downarrow \downarrow} \right], \\
P^\downarrow_z & = & \frac{2e^2/h}{G^{\uparrow \downarrow} + G^{\downarrow \downarrow}} \sum_{n^\prime,n=1}^M {\rm Im} \,
\left[{\bf t}^*_{n^\prime n, \uparrow \downarrow} {\bf t}_{n^\prime n,\downarrow \downarrow} \right].
\end{eqnarray}
\end{subequations}
Finally, if we impose the unpolarized current $n_\uparrow=n_\downarrow$ as the boundary condition in the left lead, the polarization vector of detected  current in the right lead is given by
\begin{widetext}
\begin{subequations} \label{eq:p_unpol}
\begin{eqnarray}
P^{\uparrow + \downarrow}_x  & = &   \frac{G^{\uparrow \uparrow} + G^{\uparrow \downarrow} - G^{\downarrow \uparrow} -
G^{\downarrow \downarrow}}{G^{\uparrow \uparrow} + G^{\uparrow \downarrow} + G^{\downarrow \uparrow} + G^{\downarrow \downarrow}} \label{eq:pz} \\
P^{\uparrow + \downarrow}_y & = & \frac{2e^2}{h} \frac{1}{G^{\uparrow \uparrow} + G^{\uparrow \downarrow} + G^{\downarrow \uparrow} + G^{\downarrow \downarrow}}
\sum_{n^\prime,n=1}^M {\rm Re} \, \left[ {\bf t}_{n^\prime n, \uparrow \uparrow} {\bf t}^*_{n^\prime n,\downarrow \uparrow} +  {\bf t}_{n^\prime n, \uparrow \downarrow} {\bf t}^*_{n^\prime n,\downarrow \downarrow}\right] \\
P^{\uparrow+\downarrow}_z & = & \frac{2e^2}{h} \frac{1}{G^{\uparrow \uparrow} + G^{\uparrow \downarrow} + G^{\downarrow \uparrow} + G^{\downarrow \downarrow}}
\sum_{n^\prime,n=1}^M {\rm Im} \,
\left [{\bf t}^*_{n^\prime n, \uparrow \uparrow} {\bf t}_{n^\prime n,\downarrow \uparrow} + {\bf t}^*_{n^\prime n, \uparrow \downarrow} {\bf t}_{n^\prime n,\downarrow \downarrow} \right].
\end{eqnarray}
\end{subequations}
\end{widetext}

Introducing electric~\cite{datta90,ring} or magnetic fields~\cite{popp} to manipulate spin in spintronic devices selects a preferred direction in space, thereby breaking rotational invariance. Thus, as demonstrated in Sec.~\ref{sec:ballistic} and ~\ref{sec:diffusive}, spin-resolved conductances and components of the polarization vector of the current  will depend on the direction of spin in the incoming current with respect to the direction of these fields. In the case of unpolarized injected current, all results are invariant with respect to the rotation of incoming spin since $\hat{\rho}_s  = \hat{I}_s/2$ independently of the spin quantization axis. To accommodate different polarizations of incoming current, one has to change the direction of spin quantization axis. This amounts to changing the representation of Pauli matrices Eq.~(\ref{eq:pauli}) when computing both: (i)  the transmission matrix, and (ii) polarization vector from Eq.~(\ref{eq:pol_vector}).

While the form of the spin density matrices, diagonal Pauli matrix, and the component 
of spin polarization vector $P_x^\sigma$ along the spin quantization axis are unique, 
the explicit expressions for $P_y^\sigma$ and $P_z^\sigma$ depend on particular form  
of the chosen representation for the non-diagonal Pauli matrices. The component  along the spin 
quantization axis [$P_x^\sigma$ in Eq.~(\ref{eq:pz})] has a simple physical interpretation---it 
represents normalized difference of the charge currents of spin-$\uparrow$ ($I^\uparrow = G^{\uparrow \uparrow} + G^{\uparrow \downarrow}$)  and spin-$\downarrow$ ($I^\downarrow  = G^{\downarrow \downarrow} + G^{\downarrow \uparrow}$) electrons flowing  through the right lead. The fact that our 
expression is able to reproduce commonly used Eq.~(\ref{eq:polarization}) as a 
special case demonstrates that density matrix of transported spin Eq.~(\ref{eq:rightleadunpol}) derived in Sec.~\ref{sec:density_matrix} yields rigorously defined and  unequivocal~\cite{footpareek} 
measure of spin polarization. Therefore, in the rest of the paper we reserve the term {\em 
spin polarization  of charge current}~\cite{privman,ballentine}  for $|{\bf P}|$. It is insightful 
to point out that the same spin density matrix Eq.~(\ref{eq:polarization}) also allows us to obtain 
the  vector of spin current~\cite{meso_spin_hall} $I^s = \frac{\hbar}{2e}(I^\uparrow - I^\downarrow)$, $(I^s_x, I^s_y, I^s_z) = \frac{\hbar} {2e} (P_x^\sigma I,P_y^\sigma I, P_y^\sigma I)$, flowing together 
with charge current $I = I^\uparrow + I^\downarrow =  G V$ in the  right lead of the device in Fig.~\ref{fig:setup} (biased by the voltage difference  $V$ between the leads). 

The explicit expressions for the density matrices of detected
current $\hat{\rho}^\uparrow_c$, $\hat{\rho}^\downarrow_c$,
$\hat{\rho}^{\uparrow + \downarrow}_c$, i.e.,  the corresponding
polarization vectors extracted in Eqs.~(\ref{eq:p_up})--(\ref{eq:p_unpol}), 
together with the Landauer formula for charge conductances
Eq.~(\ref{eq:conductance}), provide unified description of coupled
spin-charge transport in finite-size devices attached to external 
probes. For such structures, the system size and interfaces through 
which electrons can enter or leave the device  play an essential role 
in determining their transport properties. The proper boundary conditions, 
which require considerable effort in theoretical formalisms tailored for
infinite systems,~\cite{burkov} are intrinsically taken into
account by the Landauer-B\" uttiker scattering approach to quantum
transport. Moreover, the unified description is indispensable for
transport experiments which often detect spin current through
induced voltages on spin-selective
ferromagnetic~\cite{hammar,silsbee,jedema} or non-ferromagnetic
probes.~\cite{marcus}  The main concepts introduced here are
general enough to explain also spin-polarization in experiments
where spins are detected in optical schemes which observe the
polarization of emitted light in electroluminescence
process.~\cite{ohno}

\section{Spin coherence in transport through multichannel semiconductor nanowires} \label{sec:decoherence}

Traditional semiclassical approaches to spin
transport~\cite{jaro,dyakonov} have been focused on spin
diffusion~\cite{zhang} in disordered systems, where SO interaction
effects on transport are taken only through its role in the relaxation 
of non-equilibrium spin distribution. On the
other hand, quantum transport theories have been extensively
developed to understand the weak localization-type corrections
that SO interactions induce on the charge conduction 
properties.~\cite{larkin,lyanda,pikus} Many electrically controlled
(via SO couplings) spintronic devices  necessitate mode of operation with
ballistically propagating spin-polarized electrons (such as the original 
spin-FET proposal~\cite{datta90}) in order to
retain high degree of spin coherence. The study of spin relaxation 
dynamics in ballistic finite-size structures (such as regular or chaotic 
SO coupled quantum dots~\cite{spin_ballistic}) requires techniques 
that differ from those applied to, e.g., D'yakonov-Perel' (DP) type of 
spin relaxation in disordered systems with SO interaction 
(the DP mechanism dominates spin relaxation at low temperatures in bulk samples and 
quantum wells of III-V semiconductors). Yet another transport regime that 
requires special treatment occurs in low mobility systems whose charge
propagation is impeded by Anderson localization effects or strong
electron-phonon interactions.~\cite{spin_localized}

To quantify the degree of coherence of transported spin states in a vast range 
of transport regimes, we provide in this Section one possible implementation of 
the scattering  formalism for the spin density  matrix (Sec.~\ref{sec:density_matrix}),  
which takes as an input a microscopic Hamiltonian.  This will allow us to trace the dynamics 
of spin polarization vector of  current obtained after the injected pure spin quantum 
state propagate  through ballistic, quasiballistic, diffusive, and strongly disordered 
multichannel  semiconductor nanowires with the Rashba and/or the Dresselhaus SO couplings.

The computation of the Landauer transmission matrix ${\bf t}$ usually proceeds either phenomenologically, by replacing the device with an equivalent structure described by a 
random scattering matrix (which is applicable to specific geometries that must involve 
disorder or classical chaos due to the boundary scattering~\cite{carlo_rmt}, and 
extendable to include the SO interactions~\cite{falko}) or  by using Hamiltonian formalisms. We model semiconductor heterostructure containing a 2DEG in the $xy$-plane by an effective mass  
single-particle Hamiltonian with relevant SO interaction terms,
\begin{eqnarray} \label{eq:hamiltonian}
\hat{H} & = & \frac{\hat{p}_x^2 + \hat{p}_y^2}{2 m^*} + V_{\rm conf}(x,y) + V_{\rm disorder} (x,y) \nonumber \\
&& + \frac{\alpha}{\hbar} \left( \hat{p}_y \hat{\sigma}_x  - \hat{p}_x  \hat{\sigma}_y  \right) +
\frac{\beta}{\hbar} \left(\hat{p}_x \hat{\sigma}_x  - \hat{p}_y \hat{\sigma}_y  \right), 
\end{eqnarray}
where $m^*$ is the effective mass of an electron in semiconductor heterostructure.~\cite{footnote} Here $V_{\rm conf}(x,y)$ represents the hard-wall boundary conditions at those device  edges through which the current cannot flow. The random potential $V_{\rm disorder} (x,y)$ is zero for ballistic wires in Sec.~\ref{sec:ballistic}, and it simulates spin-independent scattering off impurities in Sec.~\ref{sec:diffusive}. In semiconductor-based devices there are two main contributions to the SO interactions: (a) electrons confined to 2DEG within  semiconductor heterostructures experience strong Rashba SO coupling [third term in Eq.~(\ref{eq:hamiltonian})] because of the low spatial symmetry of the confining potential caused by inversion-asymmetry of the space charge distribution;~\cite{rashba} (b) linear in momentum Dresselhaus SO coupling [fourth term in Eq.~(\ref{eq:hamiltonian})] which arises in semiconductors with no bulk inversion symmetry (we neglect here the cubic Dresselhaus term).~\cite{dresselhaus} In GaAs quantum well the two terms are of the same order of magnitude, while the Rashba SO coupling dominates in narrow band-gap InAs-based structures (the relative strength  $\alpha/\beta$ has recently  been extracted from photocurrent measurements~\cite{ganichev}). 

The SO coupling sets the spin precession length  $L_{\rm so} = \pi/2k_{\rm so}$ defined as 
the characteristic length scale over which spin precesses by an angle $\pi$ (i.e., the state 
$|\!\! \uparrow \rangle$ evolves into $|\!\! \downarrow \rangle$). For example, in the case of 
the Rashba SO coupling~\cite{mireles} $k_{\rm so} = m^*\alpha/\hbar^2$ ($2 k_{\rm so}$ is the 
difference of Fermi wave vectors for the spin-split  transverse energy subbands of a quantum 
wire) and~\cite{datta90}  $L_{\rm so} = \pi t_{\rm o}a/2t_{\rm so}^R$. The spin precession 
length determines evolution of spin polarization in the course of  {\em semiclassical} spatial propagation through both the ballistic~\cite{spin_ballistic} and the diffusive~\cite{dyakonov} 
SO coupled structures (which are sufficiently wide and weakly disordered, see Sec.~\ref{sec:diffusive}).

Although it is possible to evaluate the transmission matrix elements of simple systems (such as single~\cite{ring,diego} or two-channel structures~\cite{strong_rashba}) described by the  Hamiltonian Eq.~(\ref{eq:hamiltonian})  by finding the stationary states across the lead+sample systems via matching of eigenfunctions in different regions,~\cite{diego,strong_rashba,nonballistic,moroz} for efficient modeling of multichannel transport in arbitrary device geometry, as well as to include effects of disorder, it is necessary to switch  to some type of single-particle Green function technique.~\cite{datta_book} We employ here the real$\otimes$spin-space Green operators, whose evaluation requires to rewrite the Hamiltonian Eq.~(\ref{eq:hamiltonian}) in the local orbital basis
\begin{eqnarray}\label{eq:tbh}
  \hat{H} & = & \left( \sum_{\bf m} \varepsilon_{\bf m}|{\bf m} \rangle \langle {\bf m}|
  -  t_{\rm o} \sum_{\langle {\bf m},{\bf m}^\prime \rangle} |{\bf m} \rangle \langle {\bf m}^\prime| \right) \otimes \hat{I}_s \nonumber    \\ && +
  \frac{ \alpha}{\hbar} (\hat{p}_y \otimes \hat{\sigma}_x - \hat{p}_x \otimes
\hat{\sigma}_y) \nonumber \\
 && + \frac{\beta}{\hbar}(\hat{p}_x \otimes \hat{\sigma}_x - \hat{p}_y
\otimes \hat{\sigma}_y)
\end{eqnarray}
defined on the $M \times L$ lattice, where $L$ is the length of the wire in units of the lattice spacing $a$ (of the order of few nanometers when interpreted in terms of the parameters of semiconductor heterostructures employed in experiments~\cite{footnote}), and $M$ is the width of the wire. In 2D systems, $M$ is also the maximum number of conducting channels that can be opened up by positioning  $E_F$ in the band center of the Hamiltonian Eq.~(\ref{eq:tbh}). Here $t_{\rm o}=\hbar^2/(2m^*a^2)$ is the nearest-neighbor hopping between $s$-orbitals $\langle {\bf r}|{\bf m}\rangle = \psi({\bf r}-{\bf m})$ on adjacent atoms located at sites ${\bf m}=(m_x,m_y)$ of the lattice. In ballistic wires of Sec.~\ref{sec:ballistic} we set the on-site potential energy $\varepsilon_{\bf m}=0$, while the disorder in Sec.~\ref{sec:diffusive} is simulated via uniform random variable $\varepsilon_{\bf m} \in [-W/2,W/2]$. In Eq.~(\ref{eq:tbh}) $\otimes$ stand for the Kronecker product of matrices, which is the matrix representation of the tensor product of corresponding operators. The the tight-binding representation of the momentum operator is given by the matrix $\langle {\bf m} |\hat{p}_x| {\bf m}^\prime \rangle = \delta_{m_x^\prime,m_x \pm 1} i \hbar \left( m_x - m^\prime_x \right)/2a^2$. Therefore, the matrix elements of the SO terms in Eq.~(\ref{eq:tbh}) contain  spin-orbit hopping  parameters $t_{\rm so}^{\rm R}=\alpha/2a$ and $t_{\rm so}^{\rm D}=\beta/2a$, which determine the Rashba and the Dresselhaus SO coupling induced spin-splitting  of the energy bands,~\cite{mireles} respectively. All parameters in the Hamiltonian with the dimension of energy ($W$, $E_F$, $t_{\rm so}^{\rm R}$, and $t_{\rm so}^{\rm D}$) will be  expressed in Figures in the units of standard (orbital) 
hopping $t_{\rm o}=1$ of tight-binding Hamiltonians.

The spin-resolved transmission matrix elements
\begin{eqnarray}\label{eq:transmission}
  {\bf t} & = & 2 \sqrt{-\text{Im} \, \hat{\Sigma}_L^r \otimes
\hat{I}_s } \cdot \hat{G}^{r}_{1 N} \cdot
  \sqrt{-\text{Im}\, \hat{\Sigma}_R^r \otimes \hat{I}_s} \nonumber \\
    {\bf t}_{n^\prime n,\uparrow\uparrow} & \equiv & {\bf t}_{2(n^\prime-1)+1,2(n-1)+1}, \nonumber \\
    {\bf t}_{n^\prime n,\uparrow\downarrow} & \equiv & {\bf t}_{2(n^\prime-1)+1,2n}, \\
    {\bf t}_{n^\prime n,\downarrow\uparrow} & \equiv & {\bf t}_{2n^\prime,2(n-1)+1}, \nonumber \\
    {\bf t}_{n^\prime n,\downarrow\downarrow} & \equiv & {\bf t}_{2n^\prime,2n}, \nonumber
\end{eqnarray}
are obtained from the Green operator,
\begin{equation} \label{eq:green}
\hat{G}^{r}=\frac{1}{E \hat{I}_o \otimes \hat{I}_s - \hat{H}
-  \left( \begin{array}{cc}
      \hat{\Sigma}^{r}_\uparrow  &  0 \\
      0 & \hat{\Sigma}^{r}_\downarrow  \end{array} \right)},
\end{equation}
where $\hat{G}^{r}_{1 L}$ is the $2M \times 2M$ submatrix of the Green function matrix  $\hat{G}^{r}_{{\bf
mm}^\prime,{\sigma\sigma^\prime}}= \langle {\bf m},\sigma | \hat{G}^{r} |
{\bf m}^\prime, \sigma^\prime \rangle$ connecting the layers $1$ and $L$ along the direction
of transport (the $x$-axis in Fig.~\ref{fig:setup}). The Green function elements yield the probability
amplitude for an electron to propagate between two arbitrary sites (with or without flipping its spin during
the motion) inside an open conductor in the absence of inelastic processes. Here the self-energies
($r$-retarded, $a$-advanced) $\hat{\Sigma}_{L,R}^{a}=[\hat{\Sigma}_{L,R}^{r}]^{\dagger}$, $\hat{\Sigma}^{r}=\hat{\Sigma}_L^{r}+\hat{\Sigma}_R^{r}$ account  for the ``interaction''
of the open system with the left ($L$) or the right ($R$) lead.~\cite{datta_book} For simplicity,
we assume that $\hat{\Sigma}^{r}_\uparrow=\hat{\Sigma}^{r}_\downarrow$, which experimentally corresponds to identical conditions for the injection of both spin species (as realized by, e.g., two identical half-metallic ferromagnetic  leads of opposite magnetization attached to the sample~\cite{mireles}).
\begin{figure}
\centerline{\psfig{file=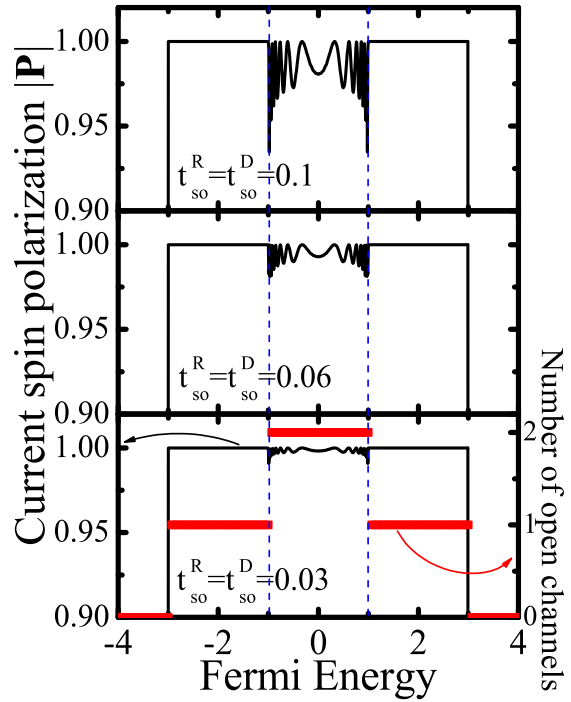,height=4.0in,width=3.0in,angle=0}
} \caption{(Color online) The degree of quantum coherence retained in spins that
have been transmitted through a clean two-channel semiconductor
nanowire, modeled on the lattice $2 \times 100$ by Hamiltonian
Eq.~(\ref{eq:tbh}), for different strengths of the Rashba and the Dresselhaus 
SO coupling tuned to $t_{\rm so}^{\rm R} = t_{\rm so}^{\rm D}$. The vertical dashed lines label the
position of the Fermi energy in the leads at which the second 
(orbital) conducting channel becomes available for injection 
and quantum transport.} \label{fig:2chbal}
\end{figure}

\subsection{Ballistic spin-charge quantum transport in semiconductor nanowires with SO interactions}\label{sec:ballistic}

Over the past two decades, a multitude of techniques has been developed  to fabricate few nanometer
wide quantum wires and explore their properties in mesoscopic transport experiments. An example is
gated two-dimensional electron gas,~\cite{vanwees} which has also become important component of 
hybrid spintronic devices.~\cite{datta90} Nevertheless, even for present nanofabrication technology 
it is still a challenge to fabricate narrow enough wires that can accommodate only one transverse 
propagating mode.

To investigate spin coherence in multichannel wires, we commence with
the simplest example---Figure~\ref{fig:2chbal} plots $|{\bf P}|$
as a function of the Fermi energy $E_F$ of electrons whose transmission
matrix ${\bf t} (E_F)$ determines spin-charge transport in quantum wire
supporting at most two ($n=1,2$) orbital conducting channels. The current
injected from the left lead is assumed to be fully polarized along
the direction of transport, as in the case of the spin-FET proposal where such setup ensures high 
level of current modulation.~\cite{tang}  As long as only one conducting 
channel is open, spin is coherent since outgoing state in the right 
lead must be of the form $ (a |\!\! \uparrow \rangle + b |\! \!
\downarrow \rangle) \otimes |n=1 \rangle$. At exactly the same
Fermi energy where the second conducting channel becomes available 
for quantum transport, the spin polarization drops below one and
spin state, therefore, loses its purity $|{\bf P}|<1$. This can be explained by 
the fact that at this $E_F$, the quantum state of transported spin of 
an electron in the right lead appears to be entangled to the ``environment" 
composed of two open orbital conducting channels of the same electron 
\begin{equation} \label{eq:schmidt}
|{\rm out} \rangle = a |\!\! \nearrow \rangle \otimes |e_1 \rangle +  b |\!\! \swarrow \rangle \otimes |e_2 \rangle.
\end{equation}
The scattering at the lead-semiconductor interface, which in the presence of the SO 
interaction give rise to the non-separable (or entangled) state in Eq.~(\ref{eq:schmidt}), 
is generated by different nature of electron states in the wire and in the leads.

\begin{figure}
\centerline{\psfig{file=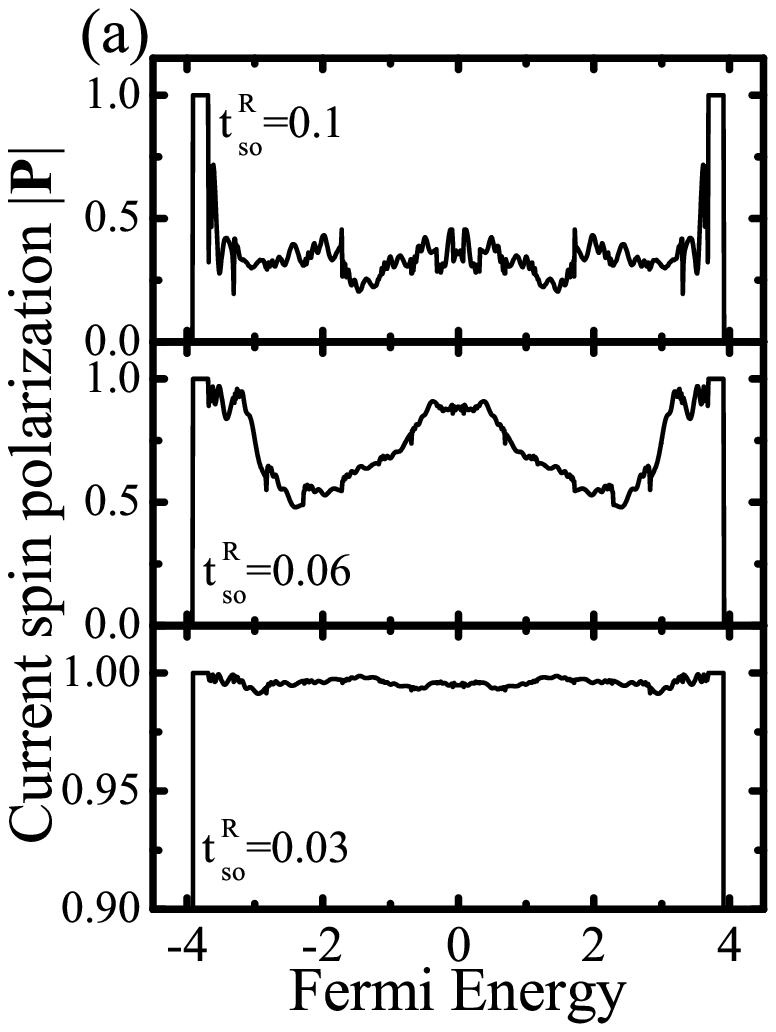,height=3.8in,width=3.0in,angle=0} }
\vspace{-0.2in}
\centerline{\psfig{file=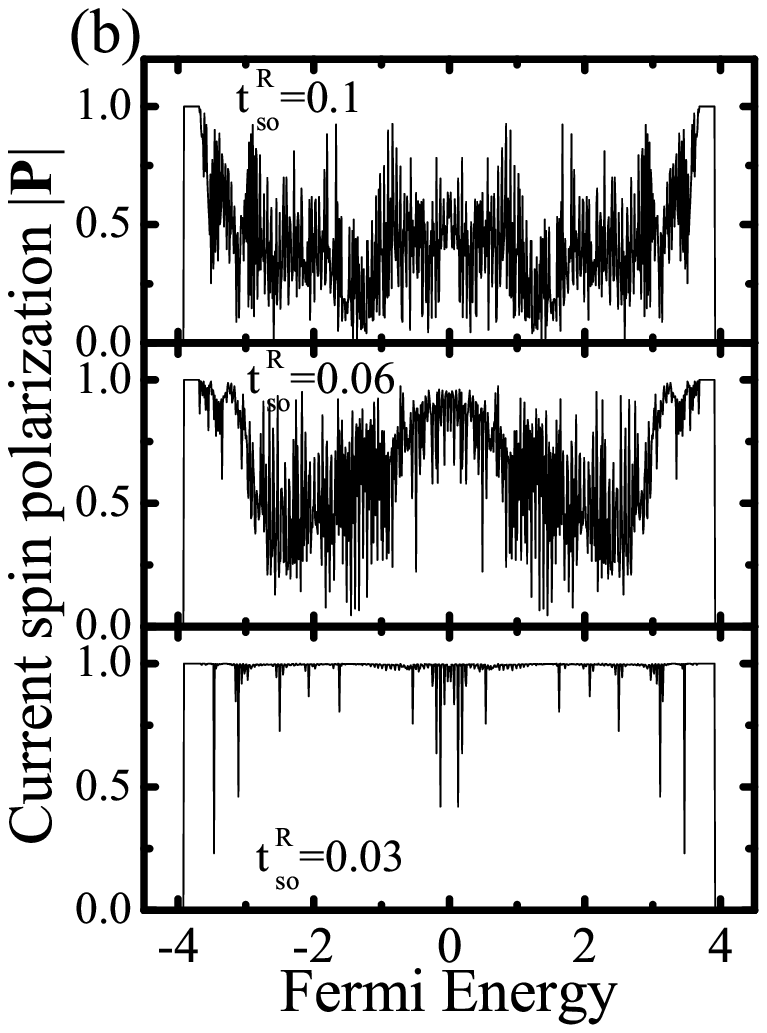,height=3.8in,width=3.0in,angle=0} }
\caption{Purity of transported spin states through a clean
semiconductor nanowire $10 \times 100$ with different strengths of the Rashba SO
coupling $t_{\rm so}^{\rm R}$. The case (a) should be contrasted with Fig.~\ref{fig:2chbal}
where the only difference is the number of transverse propagating modes (i.e., channels) in the 
leads through which electrons can be injected. In panel (b), a tunnel barrier has been
introduced between the lead and the 2DEG wire  by reducing the strength of
the lead-2DEG hopping parameter from $t_{\rm L-Sm}=t_{\rm o}$ in case
(a) to $t_{\rm L-Sm}=0.1t_{\rm o}$ in plot (b).} \label{fig:10chbal}
\end{figure}

Recent studies have pointed out that interface between ideal lead (with no SO couplings) and region 
with strong Rashba SO interaction can substantially modify spin resolved conductances~\cite{mireles} 
and suppress spin injection.~\cite{strong_rashba} Furthermore, here we unearth how moderate SO 
couplings  (the values achieved in recent experiments are of the order of~\cite{footnote} $t_{\rm so}^{\rm R} \sim 0.01$) in wires of few nanometers width will affect the coherence of ballistically transported 
spins, even when utilizing wires with Rashba=Dresselhaus SO couplings~\cite{nonballistic} (see also  Fig.~\ref{fig:nb_sfet}).  This effect becomes increasingly detrimental when more channels are opened, as demonstrated in Fig.~\ref{fig:10chbal}(a) for $M=10$ channel nanowire. Thus, such mechanism of the 
reduction of spin coherence will affect the operation of any multichannel spin-FET,~\cite{twochsfet}  independently of whether the semiconductor region is clean or disordered. Note also that injection 
through both channels of the two channel wire is not equivalent to transport with only first two 
channels  opened in the $M=10$ channel wire case because unoccupied modes can influence the transport through open channels in a way which depends on the shape of transverse  confinement potential.~\cite{hausler}

Since tunnel barriers have become an important ingredient in attempts to evade the spin injection 
impediments at the $FM$--$Sm$ interface,~\cite{injection} we introduce the tunnel barrier in the same ballistic set-up by decreasing the hoping parameter between the lead and the wire in Fig.~\ref{fig:10chbal} to $t_{\rm L-Sm}=0.1t_{\rm o}$. Although tunnel barrier inserted into an adiabatic quantum point contacts 
changes only the transmissivity of each channel without introducing the scattering between 
different channels,~\cite{nikolic_qpc} here the scattering at the interface takes place in the 
presence of SO interactions. Thus, it can  substantially affect the spin coherence of outgoing 
spins transmitted through two tunnel barriers in Fig.~\ref{fig:10chbal}(b).

To understand the transport of spin coherence along the clean wire, we plot $|{\bf P}|$  in Fig.~\ref{fig:30chbal_decay}  as a function of the wire length. Contrary to the intuition 
gained from the DP mechanism, which in unbounded diffusive systems  leads to an exponential 
decay of $|{\bf P}|$ to zero for any non-zero SO interaction, the spin coherence in clean wires 
displays oscillatory behavior along the wire or attains a residual value which exemplifies a 
partially coherent spin state. Similar behavior has been recently confirmed for 
semiclassical transport through confined disorder-free structures with integrable 
classical dynamics.~\cite{spin_ballistic} These effects depend strongly on the direction of 
spin of injected electrons with respect to the Rashba electric field (Fig.~\ref{fig:setup}) 
and on the concentration of carriers. Nevertheless, in some range of parameters apparent DP-like 
spin relaxation to zero can occur for short enough wires. This would appear as a finite spin 
coherence length in ballistic wires where no impurity scattering along the wire takes place.~\cite{cnt,spin_ballistic}
\begin{figure}
\centerline{\psfig{file=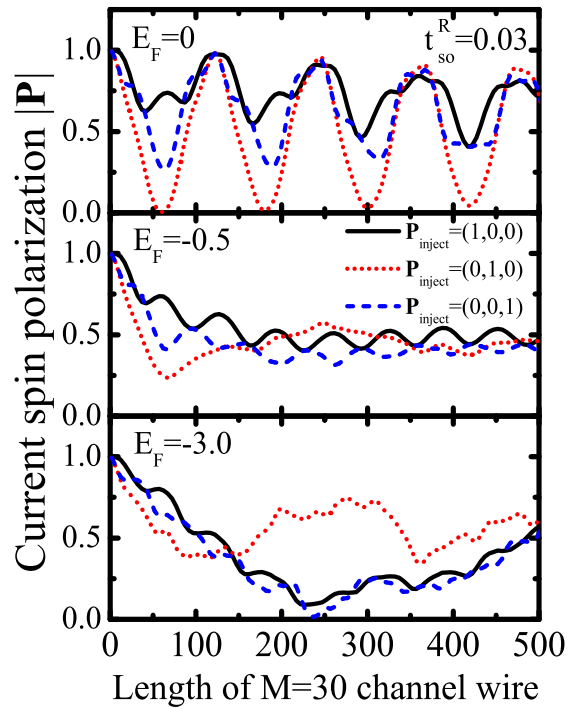,height=4.0in,width=3.0in,angle=0} }
\caption{(Color online) Transport of spin coherence along the ballistic nanowires of different length
$L$. The wires are modeled on the lattice $30 \times L$ with the Rashba SO interaction
strength $t_{\rm so}^{\rm R}=0.03$ and the corresponding spin precession length $L_{\rm so}= \pi 
t_{\rm o}a/2t_{\rm so}^{\rm R} = 52a$. The injected fully spin-polarized electron states from the 
left lead have spin-$\uparrow$ pointing in different directions with respect to the Rashba electric
field (Fig.~\ref{fig:setup}). The number of open conducting channels is: 10 at $E_F=-3.0$,
23 at $E_F=-0.5$, and 30 in the band center $E_F=0$.} \label{fig:30chbal_decay}
\end{figure}

In the absence of external magnetic fields or magnetic impurities, the SO couplings 
dominate spin dynamics in semiconductor systems with inversion asymmetry due to either 
crystalline structure or physical configuration. In such systems, they lift the spin 
degeneracy of Bloch states while at the same time enforcing a particular connection 
between wave vector and spin through the remaining Kramers degeneracy~\cite{ballentine} 
(stemming from time-reversal invariance which is not broken by the effective momentum-dependent 
magnetic field corresponding to SO interactions) of states $|{\bf k} \uparrow \rangle$ and 
$|-{\bf k} \downarrow \rangle$. For example, this leads to applied electric field inducing 
spin polarization  in addition to charge current~\cite{silsbee} or correlations between 
spin orientation and carrier velocity that is responsible for the intrinsic spin Hall effect~\cite{meso_spin_hall,murakami}. 

While coupling of spin and momentum is present in the  semiclassical transport,~\cite{silsbee,sinova_bloch} 
for quantum-coherent spatial propagation of electrons it can be, furthermore, interpreted as the 
{\em entanglement} of spinor and orbital wave function, as exemplified by the  non-separable~\cite{galindo}  quantum state in Eq.~(\ref{eq:schmidt}). Note that this type  of non-separable quantum state of describing 
a single particle has been encountered in some other situations~\cite{peres}---for example, even when the initial state is a product of a spinor and a wave function of momentum, the state transformed  by a Lorentz boost is not a direct product anymore because spin undergoes a Wigner rotation which depends on the momentum 
of the particle. These examples of  entanglement of spin and orbital degrees of freedom 
(described by state vectors belonging to two different Hilbert space) are somewhat 
different from more familiar entanglement~\cite{galindo} between different particles,  
which can be widely separated and utilized for quantum communication,~\cite{zeh,peres} because 
both degrees of freedom (spin and momentum) belong to the same particle. Nevertheless, their 
formal description proceed in the same  way---the state of the spin subsystem has to be 
described by a reduced density matrix obtained by tracing $|{\rm out} \rangle \langle {\rm out}|$ in 
Eq.~(\ref{eq:schmidt}) over the orbital degrees of freedom~\cite{zurek}
\begin{equation} \label{eq:schmidt1}
\hat{\rho}_s = {\rm Tr}_o \, |{\rm out} \rangle \langle {\rm out}| =  \left( \begin{array}{cc}
      |a|^2  &  ab^* \langle e_2 | e_1 \rangle  \\
      a^* b \langle e_1 | e_2 \rangle  & |b|^2 \end{array} \right).
\end{equation}
Here we utilize the fact that the type of quantum state in Eq.~(\ref{eq:schmidt}), containing only two terms, can  be written down for each outgoing state in the right lead for any number of open conducting channels 
$ \ge 2$. That is, such Schmidt decomposition consists of only two terms if  one of the two subsystems of a bipartite quantum system is a two-level one (independently of  how large is the Hilbert space of the other subsystem).~\cite{galindo}   

The decay of the off-diagonal elements of $\hat{\rho}_s$ in Eq.~(\ref{eq:schmidt1}), 
represented in a preferred basis ($|\!\! \uparrow \rangle$, $|\!\! \downarrow \rangle$ 
selected by the properties of incoming current), is an example of formal description of 
decoherence of quantum systems.~\cite{zeh,zurek}  The information about the superpositions  of 
spin-$\uparrow$  and spin-$\downarrow$ states is leaking into the ``environment'' 
(comprised of the orbital degrees  of freedom of one and the same electron) 
while the full quantum states still remans pure as required in mesoscopic transport. It is important to clarify that the loss of coherence in the entangled transported spin state, as an exchange of phase information between the orbital and  spin subsystems, occurs here without any energy exchange that often accompanies decoherence in solid state systems. This type of decoherence without involvement of inelastic processes can unfold at zero-temperature on the proviso that environmental quantum state is degenerate.~\cite{imry} Such situation is effectively realized in quantum transport of spin 
through multichannel wires, where full electron state remains a pure one 
$\in {\mathcal H}_o \otimes {\mathcal H}_s$  (inelastic processes would inevitably decohere 
this full state). The degeneracy of the ``environment'' here simply means that more than one 
conducting channel is open at those Fermi energies in Figs.~\ref{fig:2chbal} and ~\ref{fig:10chbal} 
where $|{\bf P}|<1$.  Note that even  when transitions between different open channels are absent (so that individual spins remain in the same channel in which they were injected and no SO entanglement takes place), the  spin density matrix of  current $\hat{\rho}_c$ can still be ``dephased''~\cite{footnote_relax,zeh} when its  off-diagonal elements are reduced due to the averaging [as in Eq.~(\ref{eq:rightleadunpol})] over states of all electrons in the detecting lead.

\subsection{Coupled spin-charge quantum diffusion in semiconductor nanowires with SO interactions} \label{sec:diffusive}

Although the problem of spin dynamics in diffusive SO coupled
semiconductors was attacked  quite some time
ago,~\cite{dyakonov} it is only recently that more involved
theoretical studies of spin-density transport in 2DEG with SO
interactions have been provoked by emerging interest in
spintronics.~\cite{burkov,wl_spin,wl_spin1,inoue} While 
standard derivations~\cite{spintronics,spin_book} of the DP spin relaxation~\cite{dyakonov} in 
semiclassical diffusive transport through bulk systems start from a 
density matrix which is diagonal in $k$-space, but allows for coherences 
in the spin Hilbert space~\cite{spintronics}, in this Section we examine 
quantum corrections to this picture in finite-size SO coupled systems 
by analyzing the decay of the off-diagonal elements of the spin density 
matrix Eq.~(\ref{eq:rightleadup}), which is obtained by tracing over 
the orbital degrees of freedom of the density matrix of pure state 
characterizing fully quantum-coherent propagation in mesoscopic systems.
\begin{figure}
\centerline{\psfig{file=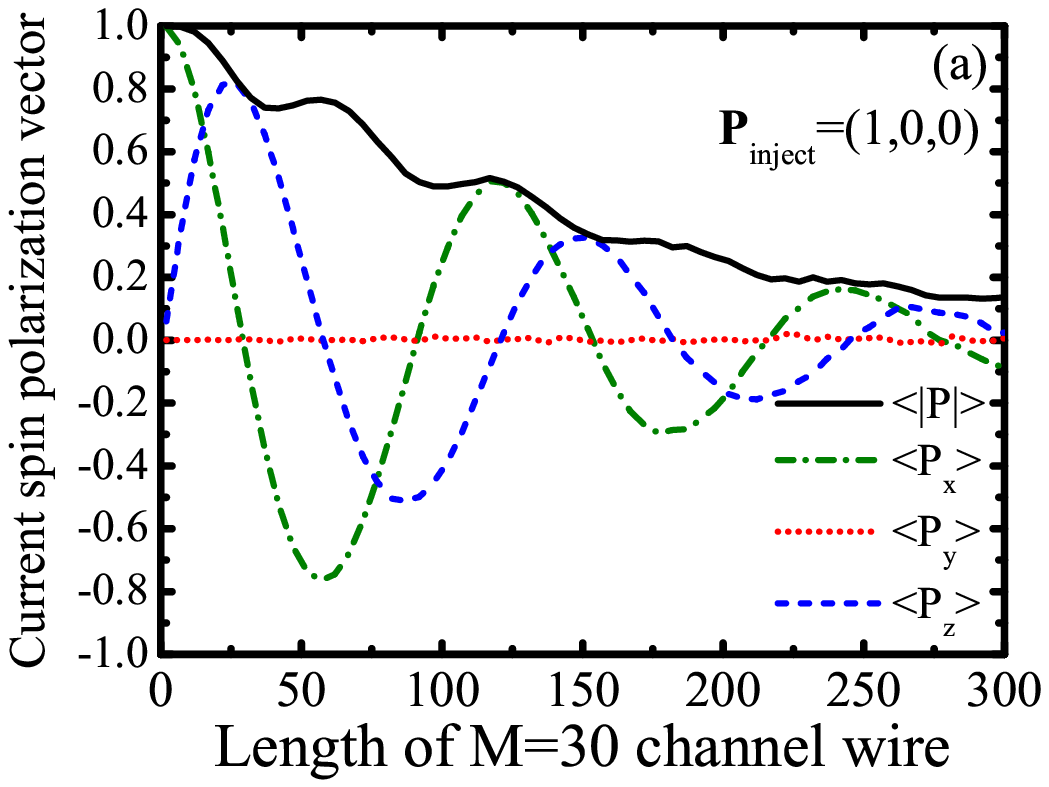,width=3.0in,angle=0} }
\centerline{\psfig{file=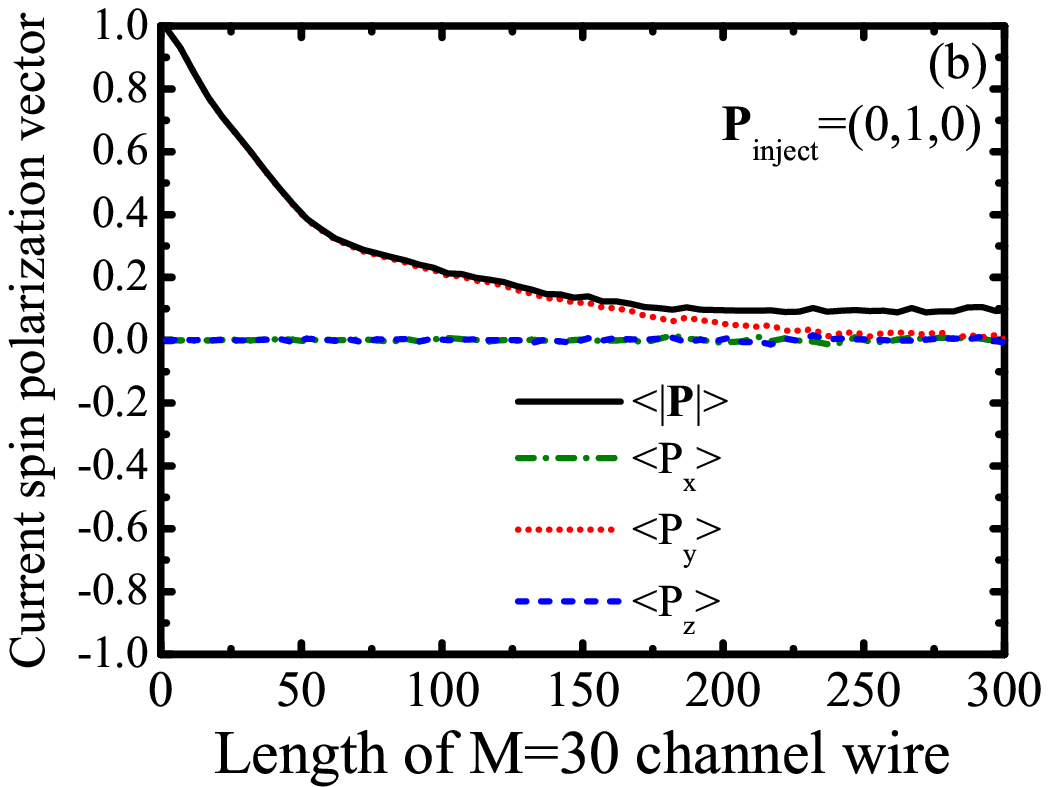,width=3.0in,angle=0} }
\centerline{\psfig{file=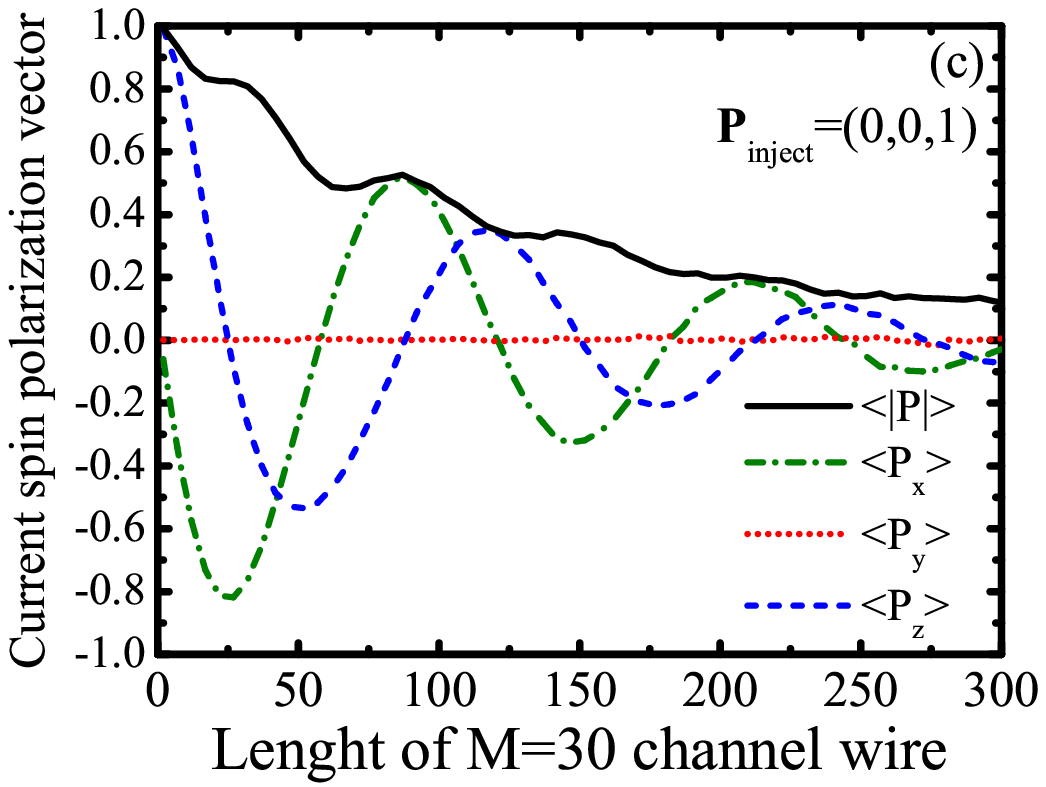,width=3.0in,angle=0} }
\caption{(Color online) The disorder-averaged components of the spin polarization
vector ($\langle P_x \rangle_{\rm dis}$, $\langle P_y \rangle_{\rm dis}$, $\langle  
P_z \rangle_{\rm dis}$), as well as its magnitude $\langle |{\bf P}| \rangle_{\rm dis}$, for the outgoing
current as a function of the length $L$ of the weakly disordered 
semiconductor  quantum wire modeled on the lattice $30 \times L$ with 
Rashba SO interaction  $t_{\rm so}^{\rm R} = 0.03$  ($L_{\rm so} = 52a$) and the 
disorder strength $W=1$ (which sets the mean free path $\ell \simeq 4a$). The injected 
electrons with $E_F=-0.5$  are spin-$\uparrow$ polarized along: (a) the $x$-axis; (b) 
the $y$-axis; and (c) the $z$-axis.} \label{fig:xyz}
\end{figure}

To facilitate comparison with our treatment of coupled spin-charge quantum transport, we 
recall here the simple semiclassical picture explaining the origin of the DP spin relaxation mechanism.~\cite{spin_ballistic} For example, if an ensemble of electrons,  
spin-polarized along the $z$-axis, is launched from the bulk of an infinite 2DEG with Rashba SO interaction $\hat{\bm{\sigma}} \cdot {\bf B}_{\rm R}({\bf k})$ in different directions, then at time $t=0$ they start to precess around the direction of the effective magnetic field ${\bf B}_{\rm R} ({\bf k})$. However, scattering off impurities and boundaries changes the direction of the electron momentum ${\bf k}$ and, therefore, can change drastically ${\bf B}_{\rm R}({\bf k})$. Averaging over an ensemble of classical trajectories leads to the decay of the $z$-component of the spin-polarization vector, whose time evolution is described by
\begin{equation} \label{eq:dp}
P_z(t) = \exp \left( -4t\ell/L^2_{\rm so} \right),
\end{equation}
assuming that spin precession length  $L_{\rm so}$ is much greater than the elastic mean 
free path $\ell = v_F \tau$.  For elastic scattering time  shorter than the
precession frequency $\tau < 1/|{\bf B}_{\rm R} ({\bf k})|$, the 
DP spin relaxation~\cite{dyakonov} is characterized by the
relaxation rate $1/\tau_s \simeq \tau {\bf B}_{\rm R}({\bf k})$.
Compared to other mechanisms of spin relaxation in semiconductors
that generate instantaneous spin flips (such as Elliot-Yafe or 
Bir-Aronov-Pikus mechanisms),~\cite{jaro} the DP spin relaxation~\cite{dyakonov} 
is a continuous process taking place during the free flight between 
scattering events. Thus, within the semiclassical framework,~\cite{jaro} 
the spin diffusion coefficient determining the relaxation of an inhomogeneous 
spin distribution is the same as the particle diffusion coefficient.
This renders the corresponding spin diffusion length $L_{\rm
sdiff} = \sqrt{{\mathcal D} \tau_s} = L_{\rm so}$ to be equal to
the ballistic spin precession length $L_{\rm so}$ and, therefore, independent of  $\ell$. The ratio $\ell/L$ controls whether the charge transport is 
diffusive ($\ell/L \ll 1$) or ballistic ($\ell/L \gg 1$).  For disordered 2DEG, modeled 
on  the 2D tight-binding lattice, the semiclassical mean free is~\cite{ando} $\ell =(6 \lambda_F^3  E_F^2)/(\pi^3 a^2 W^2)$ ($\lambda_F$ is the Fermi wavelength), which is valid for weak 
disorder $\varepsilon_{\bf m} \in [-W/2,W/2]$ in the Hamiltonian Eq.~(\ref{eq:tbh}) 
and no spin-flip scattering.

\begin{figure}
\centerline{\psfig{file=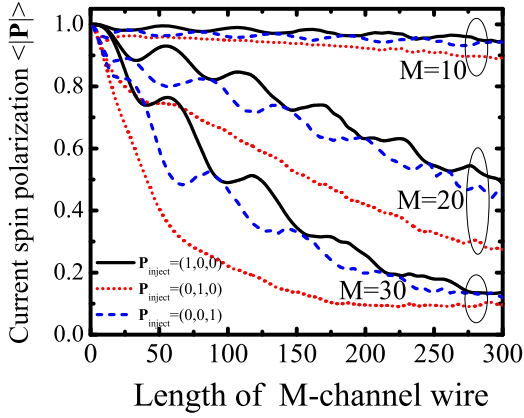,width=3.0in,angle=0} }
\caption{(Color online) The spin polarization  $\langle |{\bf P}| \rangle_{\rm dis}$ of current transmitted
through semiconductor wires of different width supporting different number of conducting
channels $M$. The nanowires are modeled on $M \times L$ lattices where quantum transport is
determined by the same set of parameters as in Fig.~\ref{fig:xyz}: $t_{\rm so}^{\rm R} = 0.03$ 
($L_{\rm so}=52a$); $W=1$ ($\ell \simeq 4a$); and $E_F=-0.5$.}\label{fig:decay_M}
\end{figure}
\begin{figure}
\centerline{\psfig{file=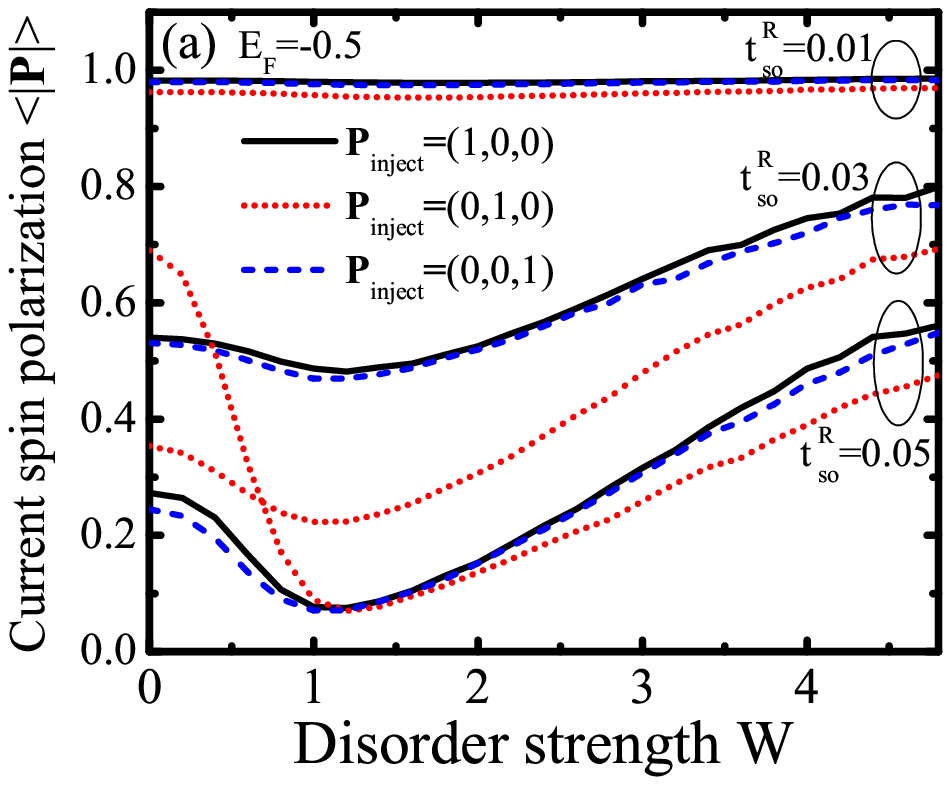,width=3.0in,angle=0} }
\centerline{\psfig{file=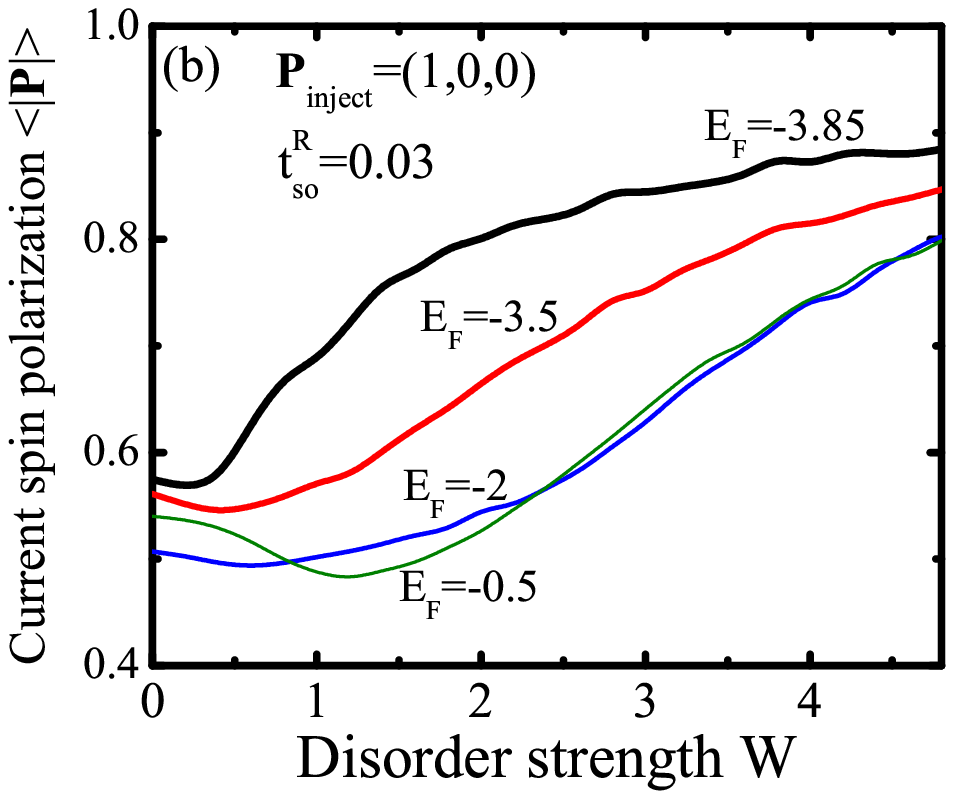,width=3.0in,angle=0} }
\caption{(Color online) The dependence of the disorder-averaged spin polarization
$\langle |{\bf P}| \rangle_{\rm dis}$ of the outgoing current, that has
been transmitted through a semiconductor quantum wire modeled on the
lattice $30 \times 100$, as a function of the disorder strength $W$ (the 
corresponding semiclassical mean free path is $\ell \simeq 16 a t_{\rm o}^2/W^2$) and 
the following parameters: (a) different values of Rashba coupling and direction of injected spin
polarization at fixed $E_F=-0.5$; (b) different Fermi energies of transported
electrons, with initial spin-$\uparrow$ polarization  along the $x$-axis, in wires
with $t_{\rm so}^{\rm R} = 0.03$ ($L_{\rm so}=52a$).} \label{fig:quantum_diffusion}
\end{figure}

To address both the fundamental issues of quantum interference corrections to spin precession and 
challenges in realization of semiconductor devices (such as the non-ballistic mode of operation~\cite{nonballistic} of the spin-FET), we introduce  the standard 
diagonal disorder $\varepsilon_{\bf m} \in [-W/2,W/2]$ in Hamiltonian Eq.~(\ref{eq:tbh}) which 
accounts for short-range isotropic spin-independent impurity potential within the wire. The principal 
spin transport quantities examined in this Section will be the disorder-averaged components of the polarization vector ($\langle P_x \rangle_{\rm dis}$, $\langle P_y \rangle_{\rm dis}$, $\langle  P_z \rangle_{\rm dis}$), as well as its magnitude $\langle |{\bf P}| \rangle_{\rm dis}$, as a function of the wire length, disorder strength $W$, and the SO coupling strengths. Note that in quasi-one-dimensional systems weak disorder can induce localization of electron states when their length $L \gg \xi$ becomes greater than the localization length $\xi=(4 M - 2) \ell$ in systems with broken spin-rotation invariance.~\cite{carlo_rmt}

In contrast to the simple exponential decay in semiclassical theory Eq.~(\ref{eq:dp}), typical 
decay of spin polarization in the multichannel quantum wire plotted in Fig.~\ref{fig:xyz} is 
more complicated. That is, the oscillatory behavior of  $\langle P_x \rangle_{\rm dis}$, $\langle P_y \rangle_{\rm dis}$, $\langle  P_z \rangle_{\rm dis}$   stems from  coherent spin precession, while the reduction of $\langle |{\bf P}| \rangle_{\rm dis}$ quantifies spin decoherence in disordered Rashba spin-split wires.  As shown in Fig.~\ref{fig:decay_M}, the decay rate of $\langle |{\bf P}| \rangle_{\rm dis}$ along the wire decreases as we decrease the wire width, thereby suppressing the DP spin relaxation 
in narrow wires.~\cite{wl_spin1} Within our quantum formalism this effect has simple interpretation---the spin decoherence is facilitated when there are many open conducting channels to which spin can entangle in 
the process of spin-independent scattering that induces transitions between the transverse subbands. In all of the phenomena analyzed here, one also has to take into account the orientation of the incoming spin with respect to the Rashba electric field in Fig.~\ref{fig:setup}. For example, when injected spin is polarized along the $y$-axis, the oscillations of polarization vector vanish  because of the fact that ${\bf B}_{\rm R}({\bf k})$ in quasi-one-dimensional systems  is nearly parallel to the direction of transverse quantization (the $y$-axis in Fig.~\ref{fig:setup}) and injected spin is, therefore, approximately an eigenstate of the Rashba Hamiltonian $\hat{\bm{\sigma}} \cdot {\bf B}_{\rm R}({\bf k})$.
\begin{figure}
\centerline{\psfig{file=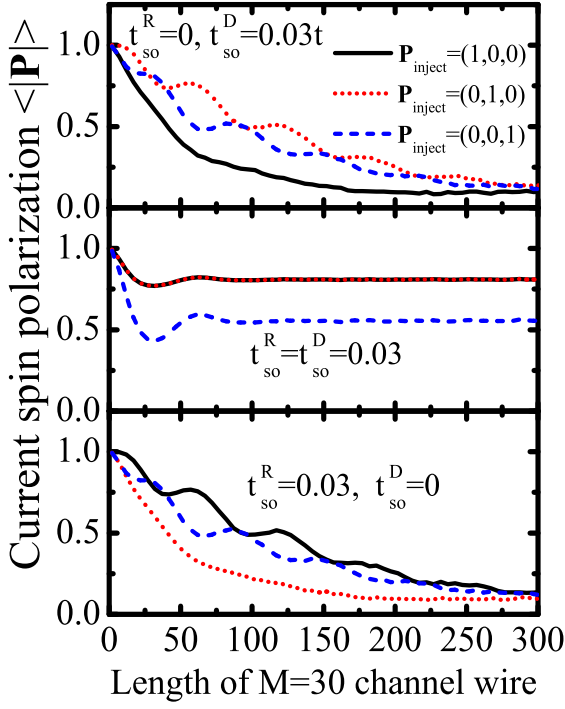,height=4.0in,width=3.0in,angle=0} }
\caption{(Color online) The degree of quantum coherence of transmitted spin states, measured by the
$\langle |{\bf P}| \rangle_{\rm dis}$, in $FMSmFM$ spin-FET-like structure with disorder
and: Dresselhaus (top panel), Rashba (bottom panel), and Rashba=Dresselhaus
(middle panel) SO couplings (as envisioned in the non-ballistic spin-FET
proposal~\cite{nonballistic}). Note that the curves for  spin-$\uparrow$ injection along
the $x$-axis and the $y$-axis overlap in the middle panel. The semiconductor region
is modeled on the lattice $30 \times L$ with disorder $W=1$ ($\ell \simeq 4a$) and $E_F=-0.5$ for
transported electrons.} \label{fig:nb_sfet}
\end{figure}
\begin{figure}
\centerline{\psfig{file=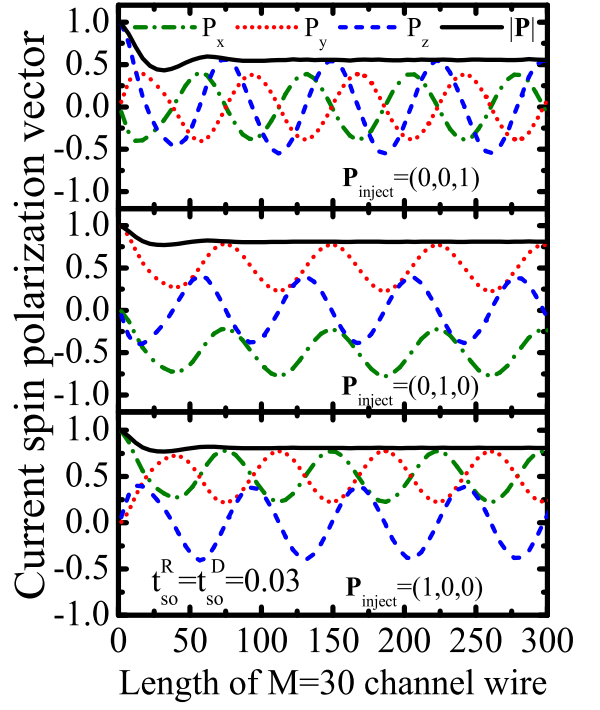,height=4.0in,width=3.0in,angle=0} }
\caption{(Color online) The components of the spin polarization vector of {\em partially
coherent} spin states that are transmitted through a non-ballistic
spin-FET~\cite{nonballistic} like structure with $t_{\rm so}^{\rm R}=t_{\rm so}^{\rm D}$. 
The structure is modeled by the same Hamiltonian used to compute the 
disorder-averaged  purity of these states $\langle |{\bf P}| \rangle_{\rm dis}$ in the middle panel of 
Fig.~\ref{fig:nb_sfet}.} \label{fig:nb_sfet1}
\end{figure}

There are salient features of $(\langle P_x \rangle_{\rm dis},\langle P_y \rangle_{\rm dis},\langle P_z \rangle_{\rm dis})$ in Fig.~\ref{fig:xyz}, brought about by SO quantum interference effects in
disordered 2DEG, that differentiate fully quantum treatment of coupled spin-charge transport from 
its semiclassical counterparts.~\cite{privman,cahay} The spin polarization $\langle |{\bf P}| \rangle_{\rm dis}$ exhibits oscillatory behavior since spin memory is preserved between successive scattering events. As the localized regime is approached, mesoscopic fluctuations of transport quantities become as large as the average value, which is therefore no longer a representative of wire properties.~\cite{carlo_rmt} For the disorder-averaged  polarization $\langle |{\bf P}| \rangle_{\rm dis}$ studied in Fig.~\ref{fig:xyz}, we notice that mesoscopic sample-to-sample fluctuations render it to be non-zero even after spin has 
traversed very long wires, i.e.,  $\left \langle \sqrt{P_x^2 + P_y^2 + P_z^2} \right \rangle_{\rm dis} \neq \sqrt{\langle P_x^2 \rangle_{\rm dis} + \langle P_y^2 \rangle_{\rm dis} + \langle P_z^2 \rangle_{\rm dis}}$.

In Fig.~\ref{fig:quantum_diffusion} we illustrate quantum
corrections to spin diffusion in {\em strongly} disordered systems,
which capture Rashba spin precession beyond the DP semiclassical~\cite{dyakonov}  
theory or weak localization corrections~\cite{wl_spin} to it (derived assuming weak 
SO coupling  in random potential that can be treated perturbatively). The 
current spin polarization $\langle |{\bf P}| \rangle_{\rm dis}$ in the wires 
of fixed length can increase with disorder even within the semiclassical regime 
$\ell > a$. This  effect survives strong Rashba interaction [panel
(a)] or opening of more channels [panel (b)]. A  conventional perturbative 
interpretation  of this effect~\cite{wl_spin,pareek,wl_spin1} is that 
quantum interference corrections to spin transport are generating longer
$\tau_s$, so that $L_{\rm sdiff}$ cease to be disorder independent. Our 
picture of spin entangled to the ``environment'' composed of orbital transport 
channels from  Sec.~\ref{sec:ballistic} sheds new light on this problem by offering 
non-perturbative  explanation for both weakly and strongly localized regime---as the 
disorder increases, some of the channels are effectively closed for transport thereby 
reducing the number of degenerate ``environmental'' quantum states that can entangle to spin.

Finally, we investigate quantum-coherence properties of spin
diffusing through multichannel wires with different types of SO
interactions. As shown in Fig.~\ref{fig:nb_sfet}, the spin
diffusion in  Rashba nanowires has the same properties as the
diffusion in the Dresselhaus ones  after one interchanges the
direction of injected polarization for situations when  incoming
spins are oriented along the $x$- and the $y$-axis. This stems from
the fact that Rashba term and linear Dresselhaus terms can be
transformed into each  other by the unitary matrix
$(\hat{\sigma}_x + \hat{\sigma}_y)/\sqrt{2}$. Therefore, the non trivial
situation arises when both of these SO interactions are present,
as shown in middle panel of Fig.~\ref{fig:nb_sfet}. 

In particular, when they are tuned to be equal $\alpha = \beta$, we find infinite
spin coherence time $L_{\rm sdiff} \rightarrow \infty$, as
discovered in the non-ballistic spin-FET
proposal.~\cite{nonballistic} However, although the current spin
polarization $\langle |{\bf P}| \rangle_{\rm dis}$ does not change
along the wire, its length-independent constant value is set below
one $\langle |{\bf P}| \rangle_{\rm dis}< 1 $ and, moreover, it is 
sensitive to the spin-polarization properties of injected current. Thus, the
transported spin in such 2DEG with carefully tuned SO couplings
will end up in a mixed quantum state which remains partially 
coherent~\cite{gefen} with constant degree of coherence along the wire. 
The partial coherence of the state is reflected in the reduced oscillations (i.e., reduced ``visibility'' 
of spin-interferences) of measurable properties  $(P_x^\sigma,P_y^\sigma,P_z^\sigma)$ 
along the nanowire, as shown in Fig.~\ref{fig:nb_sfet1} (for fully 
coherent states, where spin-$\uparrow$ and spin-$\downarrow$ interfere to form $a  |\!\!
\uparrow \rangle + b |\!\! \downarrow \rangle$, all components of
the spin polarization would oscillate between $+1$ and $-1$). While 
such states are able to evade DP spin decoherence in propagation through 
diffusive systems,~\cite{nonballistic} they are partially coherent due to 
the fact that the value of their purity is set by the scattering events at 
the lead-2DEG interface. As demonstrated by Fig.~\ref{fig:2chbal} for ballistic 
wires with Rashba=Dresselhaus couplings, the spin decoherence  processes at the  interface 
(occurring before the diffusive regime is entered) cannot be suppressed by tuning 
$\alpha = \beta$.

\section{Conclusions} \label{sec:conclusion}

We have shown how to define and evaluate the spin density matrix
of current that is transmitted through metal or semiconductor
where electrons are subjected to non-trivial spin-dependent
interactions. This formalism treats both the dynamics of spin
polarization vector and spatial propagation of charges to which
the spins are attached  in a fully quantum-coherent fashion by
employing the transmission quantities of the Landauer-B\" uttiker
scattering approach to quantum transport. Thus, it provides a
unified description of the coupled spin and charge quantum 
transport in finite-size open mesoscopic structures, while taking 
into account attached external leads and different boundary 
conditions imposed by spin injection through them.

The knowledge of the spin density matrix of electrons flowing 
through the detecting lead of a spintronic device allows us 
to quantify the degree of quantum coherence of transmitted 
spin quantum states as well as to compute the components of spin 
current flowing together with the charge current. The analysis of 
coherence properties of transported spin is {\em sine qua non} for 
the understanding of limits of all-electrical manipulation of spin via 
SO interactions in semiconductors. That  is, despite offering engineered spin
control, they can induce mechanism that lead to the decay of spin coherence, 
even in perfectly clean systems, when electrons are injected through more than one
conducting channel. We find that single spin injected through a given channel
of the left lead will end up in a partially coherent spin state in
the right lead when transitions between different transverse
subbands (due to scattering at impurities or interfaces) take
place, thereby entangling the spin quantum state to the
``environment'' composed of different orbital transverse
propagating modes. This is, therefore, a ``genuine'' decoherence
mechanism~\cite{footnote_relax,zeh} encoded in our spin density matrix. In
addition, even if every transmitted electron remains in the same
channel through which it was injected, the off-diagonal elements
of the spin density matrix  of the detected current can be reduced
(``fake'' decoherence~\cite{zeh} or ``dephasing''~\cite{footnote_relax}) due to the
averaging over different channels in multichannel transport, i.e.,
because of an incomplete description carried out by the  averaged 
density matrix~\cite{kikkawa,zeh} $\rho = 1/N \sum_{i=1}^N |\Sigma_i \rangle \langle \Sigma_i|$.

In general, reduction of  visibility of
quantum interference effects can arise due to: (i) different
phases in different transmission channels prevent conditions for
destructive  or constructive interference to be simultaneously
satisfied (even though the spin states remain fully coherent)
and/or (ii) coupling of transmitted charge or spin to other
degrees of freedom.~\cite{gefen} In the semiconductor nanowires
with different types of SO couplings studied here, each spin is
subjected to genuine decoherence mechanism via unconventional
realization of entanglement where electron spin, viewed as a
subsystems of bipartite quantum system composed of spin and
orbital degrees of freedom of a single electron, couples to open
Landauer orbital conducting channels. The 
ensemble of such spins (which are not in pure, but rather in 
improperly mixed quantum states) in the right lead is then subjected 
to ``dephasing'' when performing the averaging of their properties in 
typical transport-based spin detection schemes. Such physical
interpretation provides unified description of the decay of spin 
coherence from the ballistic to the localized transport regime.

In most of the structures examined here, the off-diagonal elements
of $\hat{\rho}_c$ do not decay completely to zero on some
characteristic time scale. Instead, in the steady state transport
through  multichannel wires with SO interaction spins will end up
in a partially coherent quantum
state.~\cite{gefen,partial_fano} The analysis of $\hat{\rho}_c$
for such states, which is  characterized by $0 < |{\bf P}| < 1$,
allows one to identify remnants of full spin interference effects,
such as the oscillations of components of spin polarization vector
shown in Fig.~\ref{fig:nb_sfet1}. The partially coherent states as
an outcome of entanglement of spin of transmitted electron with
the spin in a quantum dot have been found recently in experiments.~\cite{partial_fano} 
Here we find similar partially coherent outgoing spin states, that are,
however, induced by the physical mechanism involving entanglement
which is different and single-particle in nature. Finally, even
though current modulation through coherence dynamics of transported 
spin in spin-FET~\cite{datta90,nonballistic}  and spin-interference 
ring  devices~\cite{ring,diego,souma} will be the strongest for  single-channel 
semiconductor structures, quantum interference  effects  with partially coherent 
states could be utilized in realistic structures  that are not one-dimensional 
and not strictly ballistic.~\cite{souma}

\begin{acknowledgments}
We are grateful to S. T. Chui and J. Fabian, and L. P. Z\^ arbo for valuable 
discussions.
\end{acknowledgments}




\begin{references}
\bibitem{footnote_relax} Although vernacular terms decoherence, dephasing, 
and relaxation are often used as synonyms in spin physics, it is possible to 
make distinction between them by classifying different types of physical 
mechanisms that generate the decay of the off-diagonal elements of a two-level 
system density matrix---see the discussion on page 175 in Ref.~\onlinecite{zeh}.
\bibitem{footpareek} Our expression for the projection of spin polarization vector along the spin quantization axis Eq.~(\ref{eq:p_unpol}), which is obtained from the current spin density matrix 
derived in Eq.~(\ref{eq:rightleadunpol}), can be contrasted with some of the plausible (but hard to 
justify rigorously) measures of ``spin-polarization'' used in current literature, such as~\cite{pareek} 
$P = (G^{\uparrow \uparrow} + G^{\downarrow \downarrow} - G^{\uparrow \downarrow} - G^{\downarrow \uparrow})/(G^{\uparrow \uparrow} + G^{\downarrow \downarrow} + G^{\uparrow \downarrow} + G^{\downarrow \uparrow})$.
\bibitem{footnote} For example, InGaAs/InAlAs heterostructure fabricated in Ref.~\onlinecite{nitta} 
is characterized by  the effective mass $m^*=0.05m_0$ ($m_0$ is  the free electron mass) and the 
width of the  conduction band $\Delta=0.9$ eV, which sets $t_{\rm o}=\Delta/8=0.112$ meV for the 
orbital hopping parameter in the lattice Hamiltonian Eq.~(\ref{eq:tbh}) and $a \simeq 2.6$ nm for 
its  lattice spacing. Thus, the Rashba coupling of 2DEG formed in such heterostructures, tuned to a maximum value~\cite{nitta} $\alpha=0.93 \cdot 10^{-11}$ eVm by the gate voltage, corresponds to $t_{\rm so}^{\rm R}/t_{\rm o} \simeq 0.016$ in our model.
\end{references}
\end{document}